\documentclass[
reprint,
superscriptaddress,
preprintnumbers,
nofootinbib,
aps,
amsmath,
amssymb,
floatfix,
prl
]{revtex4-2}

\usepackage{graphicx}
\usepackage{array}
\usepackage{siunitx}
\usepackage{dcolumn}
\usepackage{bm}
\usepackage{xcolor}
\usepackage{hyperref}
\hypersetup{colorlinks=true,
citecolor=blue,
linkcolor=purple,
anchorcolor=black,
urlcolor=purple}
\usepackage{mathrsfs}
\usepackage[T1]{fontenc}
\usepackage{color} 
\usepackage{braket} 
\definecolor{red}{rgb}{0,0,0}

\newcommand{\CoFirstAuthor}{\thanks{These authors contributed equally to this work.}}

\AtBeginDocument{
  \setlength\abovedisplayskip{10pt}
  \setlength\belowdisplayskip{10pt}
}
\usepackage{etoolbox}
\usepackage{lineno}
\hypersetup{hypertexnames=false}
\makeatletter
\let\merged@author\author
\let\merged@affiliation\affiliation
\let\merged@email\email
\let\merged@thanks\thanks
\let\merged@maketitle\maketitle
\let\merged@abstract\abstract
\let\merged@endabstract\endabstract
\let\merged@addcontentsline\addcontentsline
\renewcommand\addcontentsline[3]{}
\let\auto@bib\@empty
\makeatother

\begin{document}
\title{A superconducting qutrit link beyond the qubit limit}

\author{Xiang Li}
\CoFirstAuthor
\email{francis_li@nus.edu.sg}
\affiliation{Beijing National Laboratory for Condensed Matter Physics, Institute of Physics, Chinese Academy of Sciences, Beijing 100190, China}
\affiliation{School of Physical Sciences, University of Chinese Academy of Sciences, Beijing 100049, China}
\affiliation{Centre for Quantum Technologies, National University of Singapore, 117543, Singapore}

\author{Zheng-Yang Mei}
\CoFirstAuthor
\affiliation{Beijing National Laboratory for Condensed Matter Physics, Institute of Physics, Chinese Academy of Sciences, Beijing 100190, China}
\affiliation{School of Physical Sciences, University of Chinese Academy of Sciences, Beijing 100049, China}

\author{Yang He}
\affiliation{Beijing National Laboratory for Condensed Matter Physics, Institute of Physics, Chinese Academy of Sciences, Beijing 100190, China}
\affiliation{School of Physical Sciences, University of Chinese Academy of Sciences, Beijing 100049, China}

\author{Si-Lu Zhao}
\affiliation{Beijing National Laboratory for Condensed Matter Physics, Institute of Physics, Chinese Academy of Sciences, Beijing 100190, China}
\affiliation{School of Physical Sciences, University of Chinese Academy of Sciences, Beijing 100049, China}

\author{Yan-Jun Liu}
\affiliation{Beijing National Laboratory for Condensed Matter Physics, Institute of Physics, Chinese Academy of Sciences, Beijing 100190, China}
\affiliation{School of Physical Sciences, University of Chinese Academy of Sciences, Beijing 100049, China}

\author{Xiao-Hui Song}
\affiliation{Beijing National Laboratory for Condensed Matter Physics, Institute of Physics, Chinese Academy of Sciences, Beijing 100190, China}
\affiliation{Hefei National Laboratory, Hefei 230088, China}

\author{Kai Xu}
\affiliation{Beijing National Laboratory for Condensed Matter Physics, Institute of Physics, Chinese Academy of Sciences, Beijing 100190, China}
\affiliation{Hefei National Laboratory, Hefei 230088, China}
\affiliation{Beijing Academy of Quantum Information Sciences, Beijing 100193, China}


\author{Zhong-Cheng Xiang}
\affiliation{Beijing National Laboratory for Condensed Matter Physics, Institute of Physics, Chinese Academy of Sciences, Beijing 100190, China}
\affiliation{Hefei National Laboratory, Hefei 230088, China}

\author{Dong-Ning Zheng}
\email{dzheng@iphy.ac.cn}
\affiliation{Beijing National Laboratory for Condensed Matter Physics, Institute of Physics, Chinese Academy of Sciences, Beijing 100190, China}
\affiliation{School of Physical Sciences, University of Chinese Academy of Sciences, Beijing 100049, China}
\affiliation{Hefei National Laboratory, Hefei 230088, China}

\author{Heng Fan}
\email{hfan@iphy.ac.cn}
\affiliation{Beijing National Laboratory for Condensed Matter Physics, Institute of Physics, Chinese Academy of Sciences, Beijing 100190, China}
\affiliation{School of Physical Sciences, University of Chinese Academy of Sciences, Beijing 100049, China}
\affiliation{Hefei National Laboratory, Hefei 230088, China}
\affiliation{Beijing Academy of Quantum Information Sciences, Beijing 100193, China}
\date{October 4, 2026}

\begin{abstract}
Superconducting microwave links have enabled deterministic state transfer and remote entanglement between qubits, but deterministic links have so far operated with an effectively two-dimensional transmitted Hilbert space. Here we demonstrate a superconducting qutrit link between two independently packaged nodes connected by a microwave channel. Each node combines a transmon qutrit, a transmission resonator, and a tunable Purcell-filter interface, allowing the two remote microwave-photon interfaces to be matched in both frequency and bandwidth. We implement two transition-selective photon-mediated operations that transfer the $\ket{e}$ and $\ket{f}$ qutrit components in distinct temporal modes of the same channel. We tomographically characterize arbitrary qutrit-state transfer, obtaining a mean transferred-state fidelity of $83.68\%$ and a qutrit process fidelity of $77.12\%$, exceeding both the classical qutrit-transfer benchmark and the best possible average fidelity of an effective qubit channel used to transmit an arbitrary qutrit. Using partial-transfer operations, we reconstruct a remote two-qutrit state with negativity $0.730$, a tomography-inferred dense-coding capacity of $2.273$ bits, and a tomography-inferred Collins--Gisin--Linden--Massar--Popescu (CGLMP) parameter $I_3 = 2.332$, all beyond the corresponding qubit or local bounds. These results demonstrate a superconducting microwave link that uses the native three-level structure of transmons as a genuine high-dimensional communication resource.

\end{abstract}
\maketitle

\section{Introduction}
\label{sec:1}

Coherent links between separated quantum systems are a basic ingredient of modular quantum processors and quantum networks \cite{kimble2008quantum,wehner2018quantum}. Most experimental links encode one qubit per carrier or node, but a larger local Hilbert space changes what a link can transmit. A qutrit carries a three-dimensional quantum state, and two qutrits can host Schmidt-rank-three entanglement that cannot be embedded in two qubits. Such high-dimensional resources increase dense-coding capacity, support qudit Bell inequalities and provide a larger communication alphabet \cite{bennett1992communication,horodecki2001classical,bowen2001classical,collins2002bell,erhard2020advances}. High-dimensional photonic states are well established in optics \cite{mair2001entanglement,vaziri2002experimental,cerf2002security,kues2017chip}, and local qudits have been controlled in trapped-ion and solid-state processors \cite{ringbauer2022universal,hrmo2023native,fu2022experimental}. What remains missing in superconducting microwave hardware is a deterministic remote link that transfers and entangles a full three-level state rather than one two-level subspace.

Superconducting circuits are a natural platform for this test because circuit QED combines strong artificial-atom--resonator coupling with shaped microwave emission and time-reversed capture \cite{blais2021circuit}, enabling deterministic qubit transfer and remote entanglement \cite{pechal2014microwave,wenner2014catching,kurpiers2018deterministic,axline2018demand,magnard2020microwave,burkhart2021error,zhong2021deterministic}. The higher transmon levels are native degrees of freedom, and local qutrit control and benchmarking are well developed \cite{neeley2009emulation,bianchetti2010control,xu2016coherent,morvan2021qutrit,goss2022high,brown2022trade}. A remote qutrit link, however, must map the two non-ground amplitudes onto two temporal modes of the same waveguide, preserve their relative phase and avoid opening all qutrit transitions to the continuum at once.

Here we demonstrate qutrit-level quantum communication between two independently packaged superconducting nodes connected by a one-metre microwave link. Each node contains a transmon qutrit, a transmission resonator and a tunable Purcell-filter interface. The resonator provides a transition-selective port: only the addressed qutrit--resonator sideband is opened, while the other qutrit amplitude remains stored in the transmon. The Purcell filter controls the radiative linewidth, and the transmon flux bias controls the dressed resonator frequency, allowing the two remote interfaces to be matched in situ \cite{kurpiers2019quantum}. We send the qutrit amplitude in two time bins through the same cascaded channel, using an \(e0g1\) flux-parametric sideband \cite{li2025ondemandshaped} for \(\ket{e}\) and an \(f0g1\) cavity-assisted Raman process \cite{magnard2018fast,kurpiers2018deterministic,magnard2020microwave} for \(\ket{f}\). This multimode microwave encoding is related in spirit to deterministic frequency-bin photon generation \cite{yang2025frequencybin}, but here the modes carry the non-ground amplitudes of an arbitrary qutrit and are captured by a remote node.

The results cross the qubit boundary in transfer and entanglement. Qutrit state transfer gives a mean fidelity of \(83.68\%\), above both the classical qutrit limit \(1/2\) and the maximum average fidelity \(3/4\) for transmitting an arbitrary qutrit through an effective two-dimensional channel; process tomography gives \(F_{\chi}=77.12\%\), above the corresponding bounds \(1/3\) and \(2/3\) \cite{massar1995optimal,horodecki1999general}. Partial-transfer operations generate \((\ket{gf}+\ket{ee}+\ket{fg})/\sqrt{3}\), with fidelity above the Schmidt-rank-two bound \(2/3\) and negativity \(0.730\), above the two-qubit maximum \(0.5\) \cite{terhal2000schmidt,peres1996separability,vidal2002computable,horodecki2009quantum}. The same density matrix gives a tomography-inferred dense-coding capacity of \(2.273\) bits and a tomography-inferred CGLMP parameter \(I_3=2.332\), above the relevant qubit or local bounds \cite{bennett1992communication,horodecki2001classical,bowen2001classical,collins2002bell}.

\begin{figure*}[!t]
    \centering
    \includegraphics[width=\textwidth]{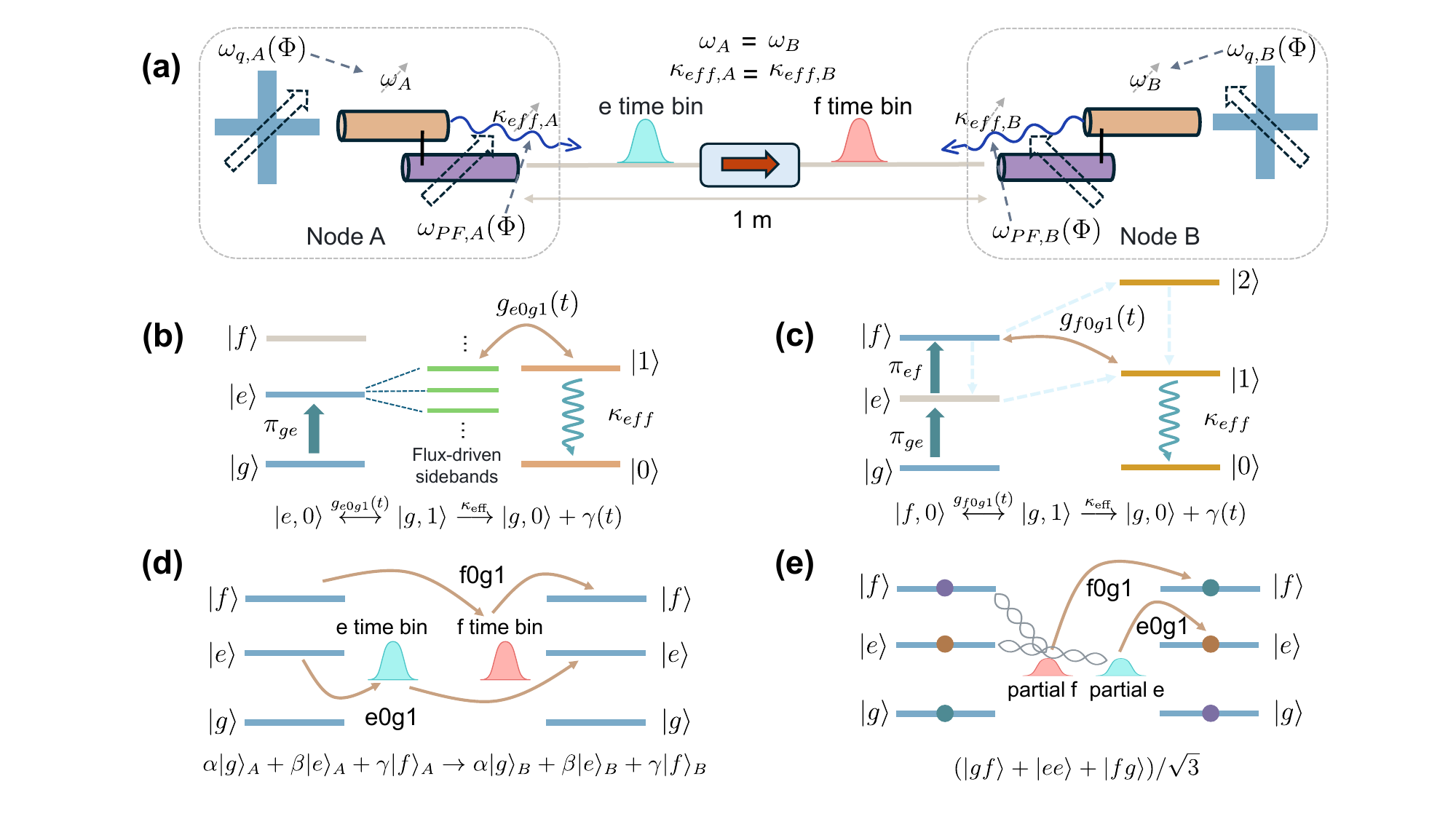}
    \caption{\textbf{Remote qutrit communication between independently packaged superconducting nodes.}
    \textbf{(a)} Two independently packaged superconducting nodes are connected by a one-metre cascaded microwave link. Each node contains a transmon qutrit, a transmission resonator and a tunable Purcell-filter interface. The transmon flux bias tunes the dressed transmission-resonator frequency, while the Purcell-filter bias tunes the radiative linewidth, allowing the two remote interfaces to be matched in both frequency, \(\omega_A=\omega_B\), and photon bandwidth, \(\kappa_{\mathrm{eff},A}=\kappa_{\mathrm{eff},B}\). The two non-ground qutrit amplitudes are carried by two temporal modes of the same link.
    \textbf{(b)} The \(e0g1\) primitive is a flux-driven sideband that couples \(|e,0\rangle\) and \(|g,1\rangle\), followed by emission through \(\kappa_{\mathrm{eff}}\).
    \textbf{(c)} The \(f0g1\) primitive uses a cavity-assisted Raman process to couple \(|f,0\rangle\) and \(|g,1\rangle\).
    \textbf{(d)} Combining the two primitives in \textbf{b} and \textbf{c} as full-transfer operations transfers an arbitrary qutrit state from node A to node B using two time bins.
    \textbf{(e)} Using the same two primitives as partial-transfer operations first entangles node A with the itinerant time-bin modes; absorption by node B maps this resource into the remote two-qutrit state \((|gf\rangle+|ee\rangle+|fg\rangle)/\sqrt{3}\).}
    \label{fig:architecture}
\end{figure*}

\section{Tunable-interface architecture and qutrit protocol}
\label{sec:architecture_protocol}

Figure~\ref{fig:architecture}(a) shows the architecture. Two superconducting nodes in separate sample boxes are connected by a one-metre microwave cable. Each node contains a transmon qutrit \cite{koch2007charge}, a transmission resonator and a tunable Purcell-filter interface to the cascaded waveguide \cite{houck2008controlling,reed2010fast,yan2023broadband}. A SQUID inserted at a current antinode of the half-wavelength Purcell-filter resonator provides flux tuning of its inductance and resonance frequency~\cite{wang2013quantum}. The transmon flux bias tunes the dressed transmission-resonator frequency, while the Purcell-filter bias tunes the radiative linewidth. These two knobs bring the remote interfaces to a common operating frequency near \(7.4347~{\rm GHz}\) and comparable photon bandwidths; the calibrations are given in Figs.~S4 and S5 of the Supplemental Material~\cite{supplemental_material}.

The qutrit channel is built from two transition-selective photon-mediated primitives, \(e0g1\) and \(f0g1\), illustrated in Fig.~\ref{fig:architecture}(b,c). The \(e0g1\) primitive couples \(\ket{e,0}\) and \(\ket{g,1}\), where the second label is the transmission-resonator photon number. A flux-parametric modulation at the qubit--resonator detuning activates this sideband \cite{didier2018analytical,li2025ondemandshaped}, allowing node A to emit the \(\ket{e}\) amplitude and node B to absorb the same temporal mode.

The \(f0g1\) primitive addresses \(\ket{f}\) through a cavity-assisted Raman process \cite{magnard2018fast,kurpiers2018deterministic,magnard2020microwave}. The Raman control couples \(\ket{f,0}\) to \(\ket{g,1}\) while the \(\ket{e}\) amplitude remains stored. The transmission resonator is therefore both the impedance-matched port and the intermediate mode that makes the port transition selective, allowing the \(\ket{e}\) and \(\ket{f}\) amplitudes to be emitted and captured in different time windows without a second physical channel.

Both primitives use chirped \(200~{\rm ns}\) sine pulses. The envelope shapes the emitted microwave field into an approximately time-symmetric temporal mode, and the chirp compensates the amplitude-dependent transition-frequency shift calibrated in Fig.~S6 of the Supplemental Material. Full-transfer operations implement the qutrit-transfer map in Fig.~\ref{fig:architecture}(d), while partial-transfer operations generate the time-bin-mediated entanglement resource in Fig.~\ref{fig:architecture}(e). The calibrated receiver-pulse delays are \(8~{\rm ns}\) for \(e0g1\) and \(13~{\rm ns}\) for \(f0g1\).

\section{Primitives of microwave qutrit transfer}
\label{sec:primitive_transfer}

\begin{figure}[t]
    \centering
    \includegraphics[width=\columnwidth]{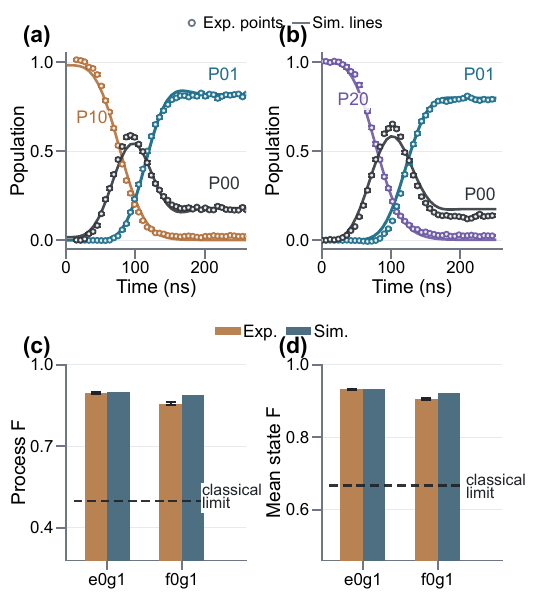}
    \caption{\textbf{Primitive transfer dynamics and two-level process benchmarks.}
    \textbf{(a,b)} Readout-corrected population dynamics for the \(e0g1\) and \(f0g1\) transfer primitives, with experimental data shown with error bars and solid curves from the pulse-level cascaded master-equation simulation.
    \textbf{(c,d)} Process-tomography benchmarks for the two primitive transfer channels. The shaded regions and dashed reference lines mark the classical two-level process-fidelity benchmark \(F_{\rm proc}=1/2\) in \textbf{c} and the corresponding mean-state-fidelity benchmark \(F_{\rm avg}=2/3\) in \textbf{d}.}
    \label{fig:primitive_transfer}
\end{figure}

We first calibrate the elementary remote links by measuring their population dynamics. In Fig.~\ref{fig:primitive_transfer}, the \(\ket{e0}\leftrightarrow\ket{g1}\) and \(\ket{f0}\leftrightarrow\ket{g1}\) transfers are plotted versus pulse time. Node A emits with the chirped sine pulse and node B applies the delayed partner pulse to absorb the same temporal mode, the microwave analogue of shaped-photon emission and time-reversed capture \cite{cirac1997quantum,yao2005theory,korotkov2011flying}. The solid curves are pulse-level cascaded master-equation simulations implemented with QuTiP \cite{gardiner1993driving,carmichael1993quantum,JOHANSSON20121760,JOHANSSON20131234}, using measured relaxation, channel efficiency, internal resonator loss and quasi-static dephasing; the calibrated couplings are listed in the Supplemental Material.

We characterize the two primitives as separate effective two-level transfer channels. The \(e0g1\) primitive transfers the \(\{\ket{g},\ket{e}\}\) component through the \(\ket{e,0}\leftrightarrow\ket{g,1}\) sideband, whereas the \(f0g1\) primitive transfers the \(\{\ket{g},\ket{f}\}\) component through the Raman-assisted \(\ket{f,0}\leftrightarrow\ket{g,1}\) channel. Two-level process tomography gives \(F_{\chi}=89.66 \pm 0.29\%\) for \(e0g1\) and \(85.68 \pm 0.60\%\) for \(f0g1\). The corresponding mean output-state fidelities, in the same order, are \(93.23 \pm 0.19\%\) and \(90.45 \pm 0.27\%\). These independently calibrated operations are the two building blocks combined below for full qutrit transfer.

\section{Arbitrary qutrit-state transfer}
\label{sec:qutrit_state_transfer}

The full qutrit-transfer sequence applies the two primitives in separate temporal windows. Node A is prepared in the qutrit state to be sent and node B starts in \(\ket{g}\). The \(f0g1\) sender pulse first maps the \(\ket{f}_A\) amplitude onto a single photon in the first temporal mode; the \(e0g1\) sender pulse then maps the \(\ket{e}_A\) amplitude onto a single photon in the second temporal mode. The \(\ket{g}_A\) component is dark to both primitives and is represented by the joint vacuum of the two time bins. The flying field therefore carries a hybrid number--time-bin qutrit,
\begin{equation}
\alpha\ket{0_1 0_2}+\gamma e^{i\varphi_f}\ket{1_1 0_2}
+\beta e^{i\varphi_e}\ket{0_1 1_2},
\end{equation}
where \(1\) and \(2\) label the \(f0g1\) and \(e0g1\) time windows, respectively. Node B applies the time-reversed capture pulses to convert the first-bin photon to \(\ket{f}_B\) and the second-bin photon to \(\ket{e}_B\). In the ideal limit,
\begin{equation}
\ket{e}_A\ket{g}_B \rightarrow \ket{g}_A\ket{e}_B,\qquad
\ket{f}_A\ket{g}_B \rightarrow \ket{g}_A\ket{f}_B,
\end{equation}
while \(\ket{g}_A\ket{g}_B\) remains dark to both transfer pulses. After the two windows, \(\alpha\ket{g}_A+\beta\ket{e}_A+\gamma\ket{f}_A\) is mapped to node B, up to deterministic phases removed with calibrated drive phases and virtual-\(Z\) gates. The pulse ordering, delay scans and phase calibrations are provided in Fig.~S7 of the Supplemental Material.

\begin{figure*}[!t]
    \centering
    \includegraphics[width=0.98\textwidth]{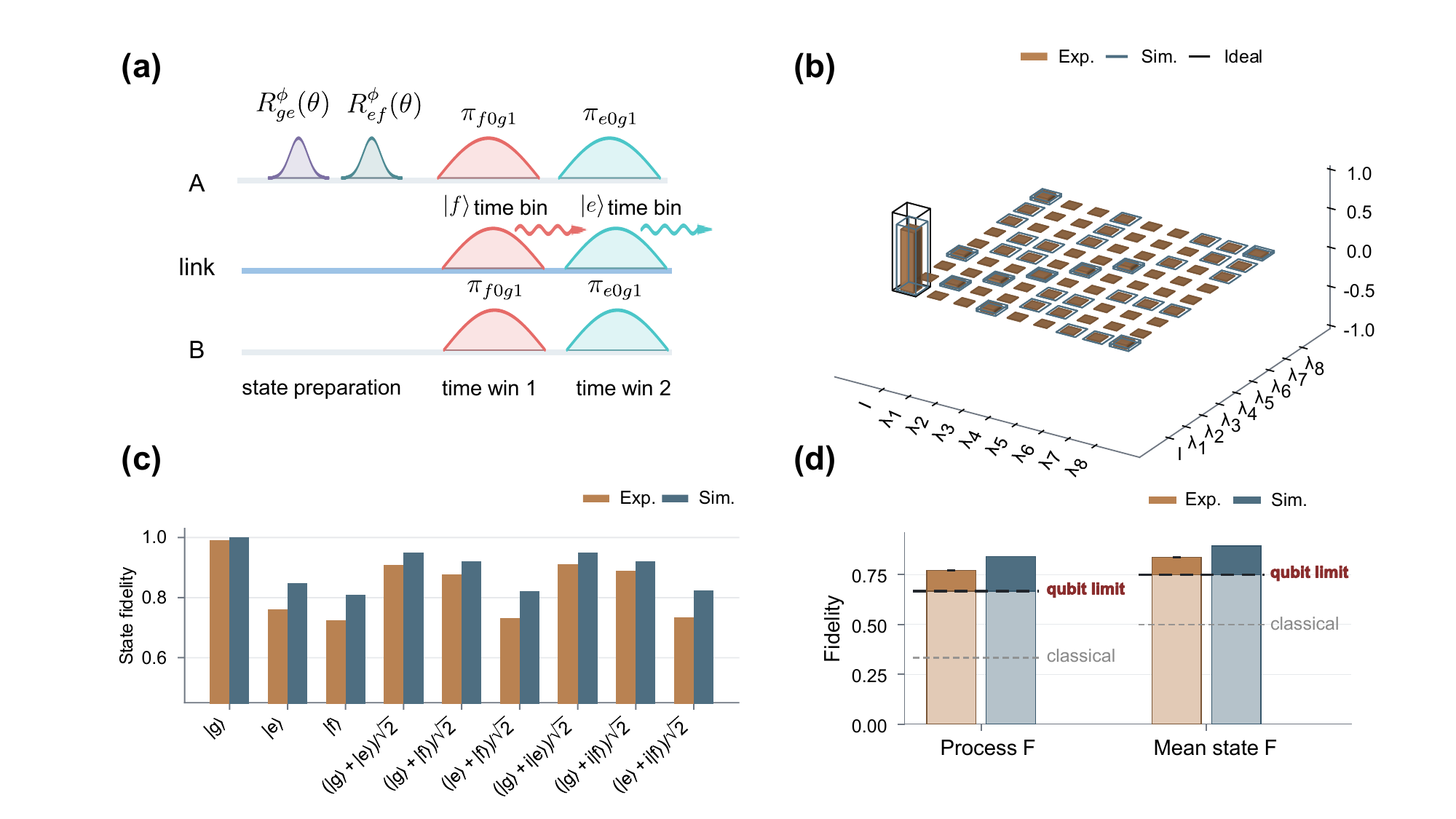}
    \caption{\textbf{Arbitrary qutrit-state transfer through the microwave link.}
    \textbf{(a)} Pulse-sequence schematic for transferring the two non-ground qutrit components with the \(f0g1\) and \(e0g1\) photon-mediated primitives. Here \(R^\phi_{ge}(\theta)\) and \(R^\phi_{ef}(\theta)\) denote rotations by angle \(\theta\) about an axis set by phase \(\phi\) in the equatorial plane of the \(ge\) and \(ef\) subspaces, respectively. In the state-transfer sequence, the \(\pi_{f0g1}\) pulse maps the \(\ket{f}\) amplitude to the first time-bin photon, and the \(\pi_{e0g1}\) pulse maps the \(\ket{e}\) amplitude to the second time-bin photon.
    \textbf{(b)} Real part of the experimental, simulated and ideal qutrit process matrix in the Gell-Mann basis. This overlaid matrix view is intended to show the dominant identity-process structure; complementary imaginary components and quantitative matrix-distance diagnostics are provided in the Supplemental Material.
    \textbf{(c)} Transferred-state fidelities for representative qutrit input states.
    \textbf{(d)} Mean state fidelity and process fidelity compared with the classical qutrit benchmark and the qubit-channel limit.}
    \label{fig:qutrit_transfer}
\end{figure*}

Qutrit state tomography uses nine analysis rotations before three-state readout, and process tomography combines nine input states with the corresponding readout-corrected output density matrices. The channel is reconstructed in the Gell-Mann basis \cite{poyatos1997complete,chuang1997prescription,nielsen2010quantum}. Figure~\ref{fig:qutrit_transfer} compares the experimental process matrix with the pulse-level simulation and the ideal identity process; complementary imaginary components and matrix-distance diagnostics are given in the Supplemental Material. The dominant identity component shows that the two time-bin primitives preserve a common qutrit phase frame.

Across the tested input states, the mean transferred-state fidelity is \(83.68 \pm 0.04\%\), above the classical qutrit measure-and-prepare limit of \(50\%\) and the maximum average fidelity \(75\%\) achievable by transmitting an arbitrary qutrit through an effective qubit channel \cite{massar1995optimal,horodecki1999general}. The measured qutrit process fidelity is \(77.12 \pm 0.10\%\). The pulse-level simulation gives \(84.10\%\) process fidelity and \(89.45\%\) mean state fidelity; the remaining differences are consistent with residual calibration errors and slow parameter drifts during the long qutrit QPT acquisition.

\section{Remote qutrit entanglement beyond the qubit limit}
\label{sec:remote_qutrit_entanglement}

\begin{figure*}[!t]
    \centering
    \includegraphics[width=0.98\textwidth]{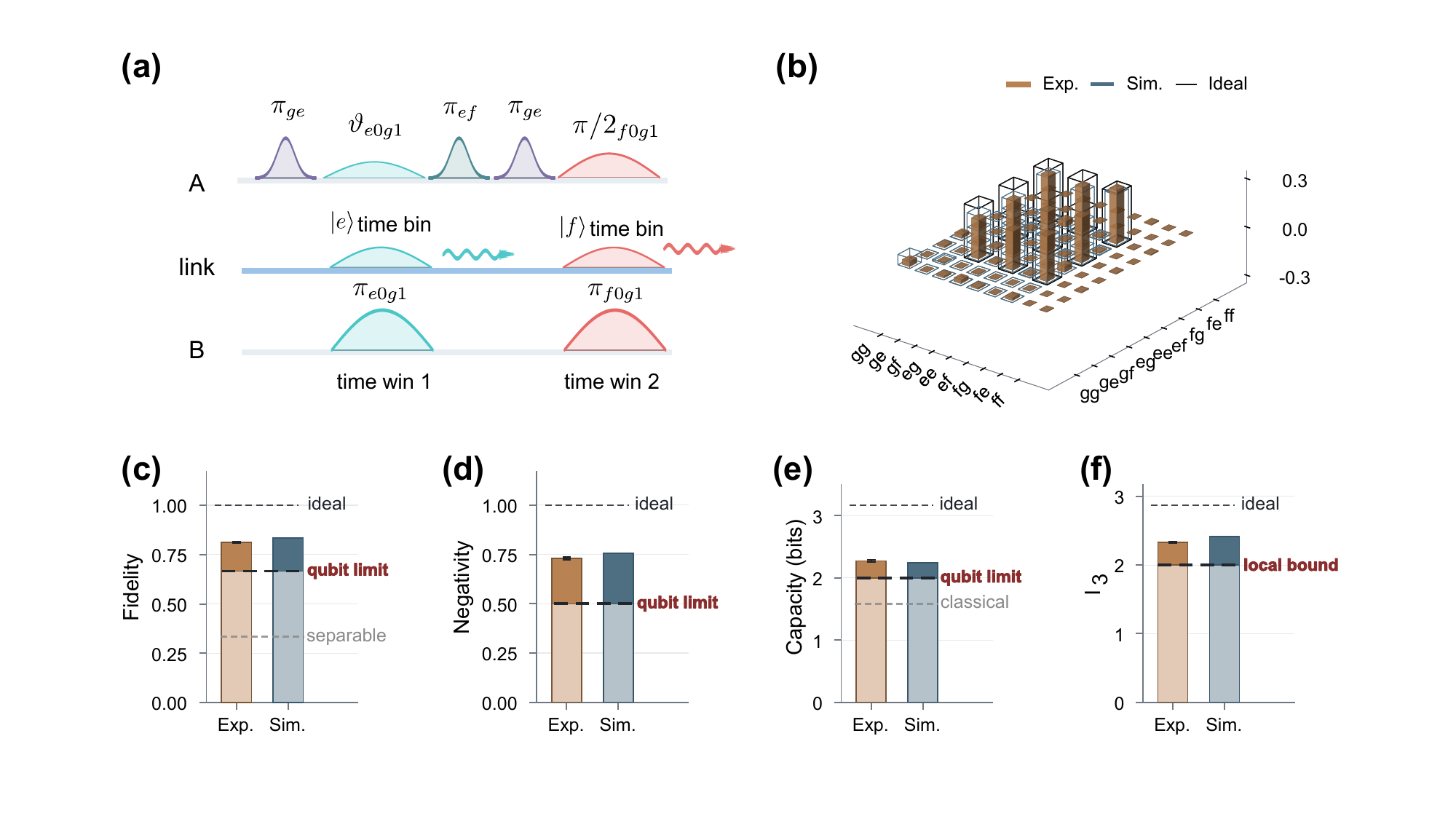}
    \caption{\textbf{Remote qutrit entanglement beyond the qubit limit.}
    \textbf{(a)} Pulse-sequence schematic for generating remote qutrit entanglement with partial \(e0g1\) and \(f0g1\) transfers. In this REG sequence, \(e0g1\) creates the first time-bin branch and \(f0g1\) creates the second time-bin branch; this ordering is distinct from the full qutrit-transfer sequence in Fig.~\ref{fig:qutrit_transfer}. The \(e0g1\) pulse angle is set to \(\vartheta_{e0g1}=2\arctan(1/\sqrt{2})\), corresponding to a target transfer probability of \(1/3\) for the \(\ket{e}\) population into the first time-bin photon. The \(f0g1\) pulse angle is set to \(\vartheta_{f0g1}=\pi/2\), corresponding to a target transfer probability of \(1/2\) for the \(\ket{f}\) population into the second time-bin photon.
    \textbf{(b)} Real part of the experimental, simulated and ideal density matrix of the generated two-qutrit state after local phase alignment. The overlaid view highlights the target populations and coherences; complementary imaginary-part, matrix-distance and readout-correction robustness diagnostics are provided in the Supplemental Material.
    \textbf{(c-f)} State fidelity, negativity, tomography-inferred dense-coding capacity and tomography-inferred CGLMP parameter compared with the relevant classical, qubit or local bounds.}
    \label{fig:qutrit_entanglement}
\end{figure*}

Remote entanglement generation uses the same photon-mediated operations before complete population exchange, but as a three-path interference protocol. Unlike qutrit transfer, the REG sequence starts with \(e0g1\): a partial operation splits the initial \(\ket{e}_A\) amplitude, local \(\pi_{ef}\) and \(\pi_{ge}\) rotations relabel the stored branches, and a subsequent partial \(f0g1\) operation creates the second time-bin branch. After full capture, the three alternatives become \(\ket{fg}\), \(\ket{ee}\) and \(\ket{gf}\), giving \((\ket{gf}+\ket{ee}+\ket{fg})/\sqrt{3}\) up to local phases. The flying state contains the vacuum, a first-bin photon or a second-bin photon, but no \(\ket{1_1 1_2}\) component, so at most one photon is exposed to propagation loss. The two-qutrit state is reconstructed from 81 joint tomography settings; its target populations and coherences dominate Fig.~\ref{fig:qutrit_entanglement}.

The state fidelity is \(81.21 \pm 0.37\%\), above the maximum \(2/3\) overlap between the target and any Schmidt-rank-two state, thereby witnessing Schmidt number 3 \cite{terhal2000schmidt}. The ensemble-averaged density matrix gives negativity \(0.730 \pm 0.005\), above the two-qubit maximum \(0.5\) \cite{peres1996separability,vidal2002computable,horodecki2009quantum}. The same reconstructed state gives a tomography-inferred dense-coding capacity of \(2.273 \pm 0.014\) bits, above both \(\log_2 3=1.585\) and the ideal qubit dense-coding limit of 2 bits, and a tomography-inferred CGLMP parameter \(I_3=2.332 \pm 0.012\), above the local bound \cite{bennett1992communication,horodecki2001classical,bowen2001classical,collins2002bell}. All REG metrics use independently calibrated readout-assignment corrections, with robustness quantified in the Supplemental Material.

\section{Discussion and Conclusion}

The experiment demonstrates coherent qutrit-state transfer and remote qutrit entanglement between separately packaged superconducting nodes. The key control element is the transmission resonator: the qutrit state is stored in the transmon, while the resonator is turned into an itinerant-photon interface only for the addressed transition. The \(e0g1\) and \(f0g1\) operations therefore convert the two non-ground amplitudes into separated temporal modes of one microwave channel, with matched bandwidths and a calibrated relative phase.

The evidence for operation beyond the qubit limit appears at three levels. The transferred-state fidelity exceeds both the classical qutrit benchmark and the best possible average fidelity of a two-dimensional channel used to transmit an arbitrary qutrit. The generated two-qutrit state has negativity above the two-qubit maximum, and its target-state fidelity witnesses Schmidt number 3. The same density matrix gives tomography-inferred dense-coding and CGLMP metrics above their qubit or local bounds. These latter quantities are calibrated-tomography resource metrics, not yet a direct dense-coding experiment or a loophole-free Bell test; they identify the state quality and measurement basis needed for those protocols.

At the current operating point the dominant limitations are finite channel efficiency, qutrit relaxation and low-frequency phase noise. The simulation error budget and forward simulations in the full cascaded transmission-resonator model show that, for \(\eta_c=0.98\), \(T_1=50~\mu{\rm s}\), Markovian \(T_\phi=20~\mu{\rm s}\), and no additional internal resonator loss, an unwindowed sech qutrit-transfer pulse gives mean state and process fidelities \(0.968\) and \(0.948\). A representative inverse-designed \(200~{\rm ns}\) REG waveform gives state fidelity \(0.939\), negativity \(0.909\), tomography-inferred dense-coding capacity \(2.718\) bits and tomography-inferred \(I_3=2.688\). The protocol therefore increases the transmitted Hilbert-space dimension without adding another physical link, and provides a route toward high-dimensional entanglement distribution in modular superconducting networks.

\clearpage

\section*{Data Availability}
The data produced in this work are available from the corresponding authors upon reasonable request.

\section*{Acknowledgments}
The authors thank Prof. Duanlu Zhou for insightful discussions. This work was supported by the Micro/Nano Fabrication Laboratory of Synergetic Extreme Condition User Facility (SECUF). Devices were made at the Nanofabrication Facilities at the Institute of Physics, CAS in Beijing. The authors thank Beijing Naishu Electronics Co., Ltd. for providing support of RF-DAC and RF-ADC based on RFSoC FPGA. This work was supported by: Quantum Science and Technology-National Science and Technology Major Project (Grant No. 2021ZD0301800), the National Natural Science Foundation of China (Grants No. 12574540, 92265207, T2121001, 12204528) 

\section*{Author Contributions}
D.-N.Z., H.F., and X.L. supervised the project. X.L. conceived the idea. X.L., Y.-J.L., and Z.-Y.M. designed the sample. X.L. performed the experiment, processed the data and wrote this manuscript. Z.-Y.M. fabricated the sample. Y.H. and S.-L.Z. helped with several iterations of the sample. All authors contributed to the discussions and production of the manuscript. 

\section*{Competing Interests}
The authors declare no competing interests.

\nocite{bronn2015broadband,caves1982quantum,Silva2010schemes,srinivasan2014time}
\makeatletter\let\pre@bibdata\@empty\makeatother
%

\onecolumngrid
\thispagestyle{article}
\clearpage
\makeatletter
\do@output@cclv{}
\@booleantrue\preprintsty@sw
\@booleanfalse\twocolumn@sw
\@twocolumnfalse
\def\@pointsize{12}
\let\merged@setup@hook\setup@hook
\let\setup@hook\@empty
\input{aps12pt4-2.rtx}
\setup@hook
\let\setup@hook\merged@setup@hook
\tolerance200
\hfuzz0.1pt
\vfuzz0.1pt
\textheight665.5pt
\set@column@hsize\@ne
\@colht\textheight
\@colroom\textheight
\vsize\textheight
\setlength\abovecaptionskip{10pt}
\setlength\belowcaptionskip{2pt}
\ps@preprint
\def\frontmatter@abstractwidth{\textwidth}
\def\frontmatter@affiliationfont{\it}
\let\section\section@preprintsty
\let\subsection\subsection@preprintsty
\let\subsubsection\subsubsection@preprintsty
\set@page@grid
\let\set@footnotewidth\set@footnotewidth@one
\let\compose@footnotes\compose@footnotes@one
\def\title@column#1{\minipagefootnote@init#1\minipagefootnote@foot}
\def\close@column{\newpage}
\let\author\merged@author
\let\affiliation\merged@affiliation
\let\email\merged@email
\let\thanks\merged@thanks
\let\maketitle\merged@maketitle
\let\abstract\merged@abstract
\let\endabstract\merged@endabstract
\let\addcontentsline\merged@addcontentsline
\frontmatter@init
\global\setbox\absbox\box\voidb@x
\setcounter{section}{0}
\setcounter{subsection}{0}
\setcounter{subsubsection}{0}
\setcounter{footnote}{0}
\setcounter{mpfootnote}{0}
\patchcmd{\titleblock@produce}{\label{FirstPage}}{\label{SuppFirstPage}}{}{}
\patchcmd{\frontmatter@footnotemark}{frontmatter.}{supp.frontmatter.}{}{}
\patchcmd{\frontmatter@makefntext}{frontmatter.}{supp.frontmatter.}{}{}
\patchcmd{\present@bibnote}{frontmatter.}{supp.frontmatter.}{}{}
\def\NAT@bibsetnum#1{\setlength{\topsep}{\z@}\NATx@bibsetnum{\ref{SuppLastBibItem}}}
\patchcmd{\endthebibliography}{LastBibItem}{SuppLastBibItem}{}{\ClassError{revtex4-2}{Unable to separate the supplementary bibliography labels}{}}
\makeatother
\normalfont\normalsize


\title{Supplemental Material for ``A superconducting qutrit link beyond the qubit limit''}

\author{Xiang Li}
\email{francis_li@nus.edu.sg}
\affiliation{Beijing National Laboratory for Condensed Matter Physics, Institute of Physics, Chinese Academy of Sciences, Beijing 100190, China}
\affiliation{School of Physical Sciences, University of Chinese Academy of Sciences, Beijing 100049, China}
\affiliation{Centre for Quantum Technologies, National University of Singapore, 117543, Singapore}

\author{Zheng-Yang Mei}
\affiliation{Beijing National Laboratory for Condensed Matter Physics, Institute of Physics, Chinese Academy of Sciences, Beijing 100190, China}
\affiliation{School of Physical Sciences, University of Chinese Academy of Sciences, Beijing 100049, China}

\author{Yang He}
\affiliation{Beijing National Laboratory for Condensed Matter Physics, Institute of Physics, Chinese Academy of Sciences, Beijing 100190, China}
\affiliation{School of Physical Sciences, University of Chinese Academy of Sciences, Beijing 100049, China}

\author{Si-Lu Zhao}
\affiliation{Beijing National Laboratory for Condensed Matter Physics, Institute of Physics, Chinese Academy of Sciences, Beijing 100190, China}
\affiliation{School of Physical Sciences, University of Chinese Academy of Sciences, Beijing 100049, China}

\author{Yan-Jun Liu}
\affiliation{Beijing National Laboratory for Condensed Matter Physics, Institute of Physics, Chinese Academy of Sciences, Beijing 100190, China}
\affiliation{School of Physical Sciences, University of Chinese Academy of Sciences, Beijing 100049, China}

\author{Xiao-Hui Song}
\affiliation{Beijing National Laboratory for Condensed Matter Physics, Institute of Physics, Chinese Academy of Sciences, Beijing 100190, China}
\affiliation{Hefei National Laboratory, Hefei 230088, China}

\author{Kai Xu}
\affiliation{Beijing National Laboratory for Condensed Matter Physics, Institute of Physics, Chinese Academy of Sciences, Beijing 100190, China}
\affiliation{Hefei National Laboratory, Hefei 230088, China}
\affiliation{Beijing Academy of Quantum Information Sciences, Beijing 100193, China}


\author{Zhong-Cheng Xiang}
\affiliation{Beijing National Laboratory for Condensed Matter Physics, Institute of Physics, Chinese Academy of Sciences, Beijing 100190, China}
\affiliation{Hefei National Laboratory, Hefei 230088, China}

\author{Dong-Ning Zheng}
\email{dzheng@iphy.ac.cn}
\affiliation{Beijing National Laboratory for Condensed Matter Physics, Institute of Physics, Chinese Academy of Sciences, Beijing 100190, China}
\affiliation{School of Physical Sciences, University of Chinese Academy of Sciences, Beijing 100049, China}
\affiliation{Hefei National Laboratory, Hefei 230088, China}

\author{Heng Fan}
\email{hfan@iphy.ac.cn}
\affiliation{Beijing National Laboratory for Condensed Matter Physics, Institute of Physics, Chinese Academy of Sciences, Beijing 100190, China}
\affiliation{School of Physical Sciences, University of Chinese Academy of Sciences, Beijing 100049, China}
\affiliation{Hefei National Laboratory, Hefei 230088, China}
\affiliation{Beijing Academy of Quantum Information Sciences, Beijing 100193, China}
\date{October 4, 2026}
\maketitle

\setcounter{equation}{0}
\setcounter{figure}{0}
\setcounter{table}{0}
\setcounter{page}{1}

\renewcommand{\theequation}{S\arabic{equation}}
\renewcommand{\thefigure}{S\arabic{figure}}
\renewcommand{\thetable}{S\arabic{table}}

\setcounter{tocdepth}{1}

\newcommand{\suppfiginclude}[2]{%
  \IfFileExists{#1}{%
    \includegraphics[#2]{#1}%
  }{%
    \fbox{%
      \begin{minipage}[c][0.22\textheight][c]{0.86\linewidth}
      \centering
      Missing figure file: \texttt{\detokenize{#1}}
      \end{minipage}%
    }%
  }%
}

\clearpage
\tableofcontents
\clearpage

\noindent\textbf{Supplementary roadmap.}
The table below summarizes how the Supplemental Material supports the main-text claims and figures.
\begin{center}
\footnotesize
\begin{minipage}{0.92\linewidth}
\hrule
\vspace{0.35em}
\noindent
\begin{minipage}[t]{0.32\linewidth}
\raggedright\textbf{Main-text claim or figure}
\end{minipage}\hfill
\begin{minipage}[t]{0.63\linewidth}
\raggedright\textbf{Supporting Supplementary material and key diagnostics}
\end{minipage}
\vspace{0.35em}
\hrule
\vspace{0.45em}
\noindent
\begin{minipage}[t]{0.32\linewidth}
\raggedright Matched qutrit-link architecture and photon interface, Fig.~1
\end{minipage}\hfill
\begin{minipage}[t]{0.63\linewidth}
\raggedright Sec.~I and Figs.~S1--S6: device layout, wiring, operating points, frequency matching, tunable linewidth calibration and chirped-pulse calibration.
\end{minipage}
\par\vspace{0.45em}
\noindent
\begin{minipage}[t]{0.32\linewidth}
\raggedright Primitive transfer and arbitrary qutrit-state transfer, Figs.~2 and 3
\end{minipage}\hfill
\begin{minipage}[t]{0.63\linewidth}
\raggedright Sec.~II, Figs.~S7--S10 and Table~S2: delay and phase calibration, logical state maps, tomography bases, independently calibrated readout correction, imaginary components and matrix-distance diagnostics.
\end{minipage}
\par\vspace{0.45em}
\noindent
\begin{minipage}[t]{0.32\linewidth}
\raggedright Remote qutrit entanglement and beyond-qubit benchmarks, Fig.~4
\end{minipage}\hfill
\begin{minipage}[t]{0.63\linewidth}
\raggedright Sec.~III and Fig.~S11: metric definitions, readout-correction sensitivity, assignment-matrix resampling and repetition-level confidence checks for fidelity, negativity, dense-coding capacity and \(I_3\).
\end{minipage}
\par\vspace{0.45em}
\noindent
\begin{minipage}[t]{0.32\linewidth}
\raggedright Pulse-level simulation, waveform modelling and improvement priorities
\end{minipage}\hfill
\begin{minipage}[t]{0.63\linewidth}
\raggedright Sec.~IV, Figs.~S12--S14 and Tables~S3--S5: cascaded master-equation model, quasi-static dephasing implementation, sine/sech waveform comparison, inverse-designed REG waveform construction and cumulative error budget.
\end{minipage}
\par\vspace{0.45em}
\hrule
\end{minipage}
\end{center}

\section{\MakeUppercase{Device, wiring and matched photon interface}}
\label{sec:supp_hardware_setup}
\label{sec:supp_device_architecture_calibrated_operating_point}
\label{sec:supp_interface_matching_pulse_calibration}

\subsection{Sample layout and node architecture}
\label{sec:supp_sample_layout}

\begin{figure}[htbp]
    \centering
    \suppfiginclude{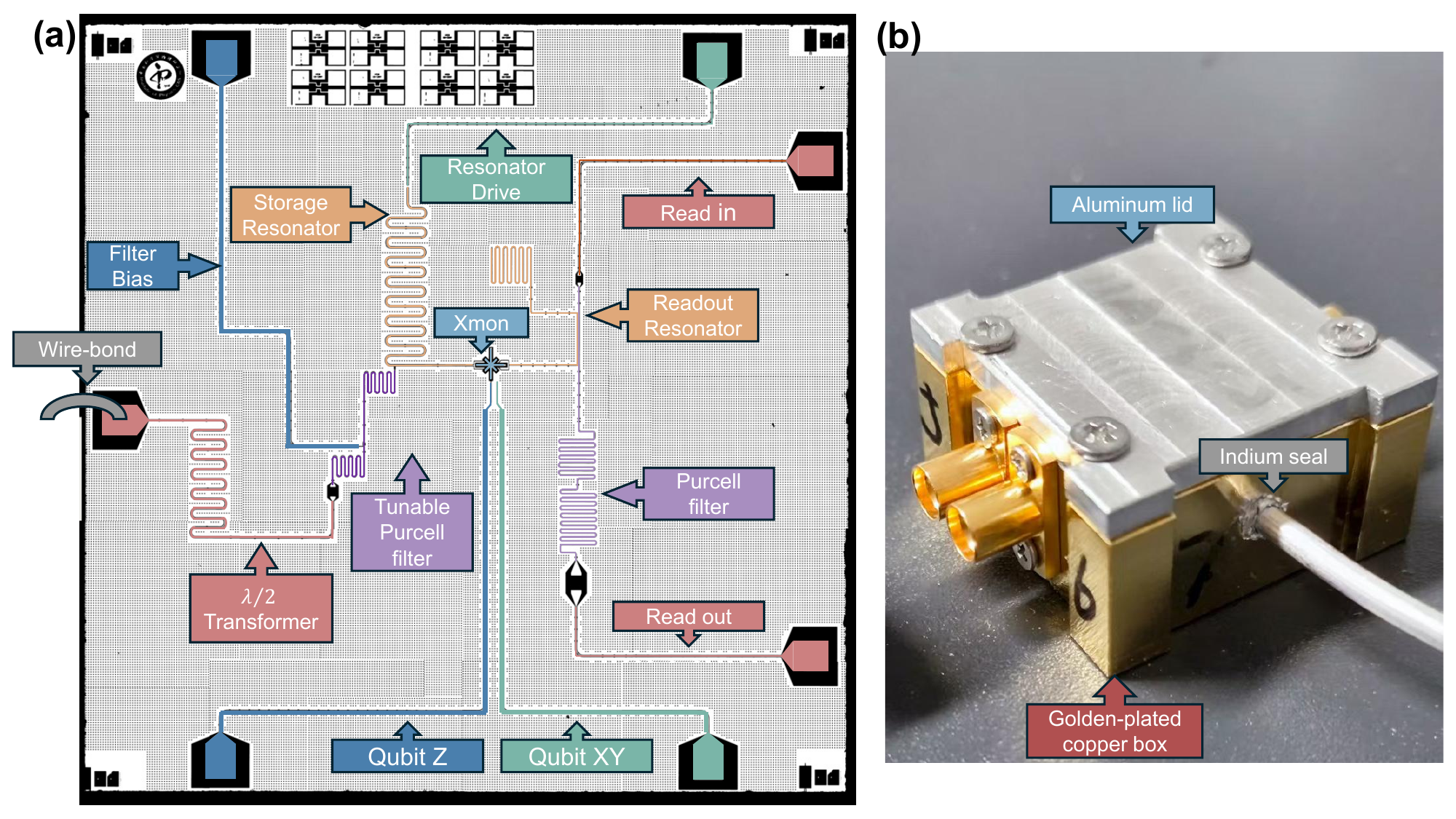}{width=0.95\textwidth}
    \caption{\textbf{Sample layout and packaged two-node device.}
    \textbf{(a)} Device structure of one superconducting qutrit node. Each node contains a flux-tunable asymmetric-junction transmon, a transmission resonator used for photon emission and capture, a readout resonator with its own Purcell filter, and a tunable Purcell-filter interface to the microwave link.
    \textbf{(b)} Packaged sample box after assembly. The package uses an aluminium lid, a gold-plated copper box and indium seals around the aluminium coaxial-cable ports to shield the chip from residual free-space infrared radiation inside the dilution refrigerator. The two independently packaged nodes are connected by the directional microwave link used for cascaded state transfer and remote entanglement generation.}
    \label{fig:supp_sample_layout}
\end{figure}

Figure~\ref{fig:supp_sample_layout} shows the physical implementation of the two-node link. Each node contains a flux-tunable asymmetric-junction transmon used as a qutrit, a transmission resonator used for photon emission and capture, a readout resonator with a Purcell filter, and a tunable Purcell-filter interface to the microwave link. The tunable interface is connected to the bonding pad through a half-wavelength impedance transformer. This transformer places a distributed-current node near the wire-bond position, reducing Joule heating and suppressing additional loss introduced by the wire bonds in the inter-node link. The chip is enclosed in a gold-plated copper box with an aluminium lid, and indium seals are used around the aluminium coaxial-cable ports. This package reduces direct infrared exposure of the chip to residual free-space radiation in the dilution refrigerator. The two chips are mounted in independent sample boxes and connected through a directional microwave link.

\subsection{Cryogenic wiring and measurement setup}
\label{sec:supp_measurement_chain}

\begin{figure}[htbp]
    \centering
    \suppfiginclude{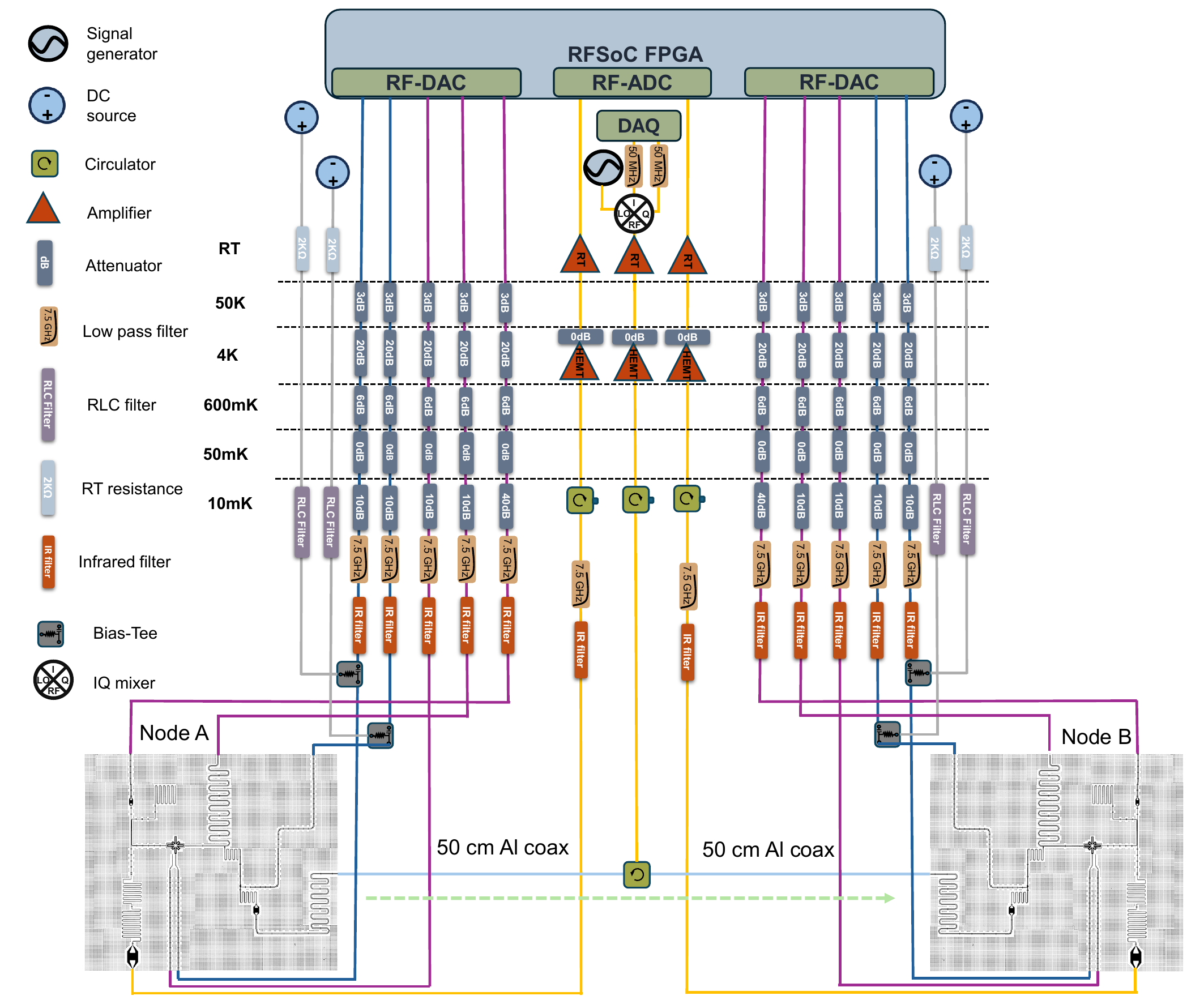}{width=0.95\textwidth}
    \caption{\textbf{Cryogenic wiring and room-temperature control setup.}
    The diagram shows the cryogenic wiring, room-temperature microwave-control and digitization chain, and the directional cascaded-channel layout used for photon emission from node A and capture by node B. It includes the input attenuation and filtering, qubit microwave-control lines, qutrit flux-modulation lines, Purcell-filter \(z\)-control lines, storage-resonator drive lines, readout lines, circulators/isolators, amplification chain, RFSoC/RF-DAC/RF-ADC hardware, mixers, local oscillators, and the one-metre inter-node link.}
    \label{fig:supp_measurement_chain}
\end{figure}

Figure~\ref{fig:supp_measurement_chain} summarizes the wiring and control stack used for the experiment. The samples are mounted at the mixing-chamber stage of a Bluefors dilution refrigerator. The microwave input lines contain cryogenic attenuation and infrared/low-pass filtering before reaching the sample boxes. The total attenuation is \(39~{\rm dB}\) on both the qutrit flux-control (\(Z\)) lines and the microwave-control (\(XY\)) lines, and \(69~{\rm dB}\) on the readout input line. Local qutrit rotations are applied through the qubit microwave-control lines. The \(e0g1\) primitive is driven through the qutrit flux-modulation line, while the \(f0g1\) Raman pulse is applied through the microwave-control chain. Compared with the setup used in our earlier shaped-photon experiment~\cite{supp:li2025ondemandshaped}, each node also includes an additional Purcell-filter \(z\)-control line, which tunes the coupling between the transmission resonator and the waveguide. The storage-resonator drive lines are wired for resonator characterization and future direct resonator operations; they are not used in the qutrit-link pulse sequences analyzed in the main text.

The room-temperature electronics follow the same RFSoC-based architecture as in our previous photon-shaping experiment~\cite{supp:li2025ondemandshaped}. The control pulses are generated by RF-DAC channels implemented on Xilinx Zynq UltraScale+ RFSoC XCZU47DR boards, operated at a sampling rate of \(8.0~{\rm GSa/s}\). These channels generate the qutrit-control pulses, readout pulses, flux-parametric modulation pulses, Raman pulses and Purcell-filter \(z\)-pulses, with additional room-temperature filtering, amplification and IQ mixing used where required by the frequency band. The readout signals are routed through separate readout lines with cryogenic non-reciprocal elements and a low-noise amplification chain before room-temperature demodulation and digitization by RF-ADC channels based on Xilinx Zynq UltraScale+ RFSoC XCZU47DR boards operated at \(4.0~{\rm GSa/s}\). Local oscillators and the RFSoC clocks are referenced to a rubidium frequency standard (Stanford Research Systems FS725), providing phase-coherent up-conversion, down-conversion and digitization during each acquisition block. The photon-transfer path is directional and separate from the readout chains: photons emitted from the transmission resonator of node A propagate through the one-metre inter-node link and are captured by the matched transmission resonator of node B. The microwave and flux-control channels are synchronized by the room-temperature control electronics. The calibrated relative delays reported in the main text and in Fig.~\ref{fig:supp_primitive_transfer_alternate} therefore include both microwave propagation through the link and fixed cable/electronics delays.

\subsection{Device parameters and operating points}
\label{sec:supp_device_parameters}

\begin{table}[htbp]
\caption{\textbf{Device parameters and operating points.}
The table lists independently calibrated operating-point parameters used in the qutrit-link experiment and in the pulse-level simulations.}
\label{tab:supp_device_parameters}
\centering
\begin{tabular}{lcc}
\hline
Parameter & Node A & Node B \\
\hline
Qutrit transition frequency \(\omega_{ge}/2\pi\) & \(6131.1~{\rm MHz}\) & \(6131.8~{\rm MHz}\) \\
Qutrit transition frequency \(\omega_{ef}/2\pi\) & \(5921.4~{\rm MHz}\) & \(5922.6~{\rm MHz}\) \\
Anharmonicity \(\alpha/2\pi\) & \(-209.8~{\rm MHz}\) & \(-209.1~{\rm MHz}\) \\
Auxiliary-resonator frequency \(\omega_{\rm aux}/2\pi\) & \(7.4347~{\rm GHz}\) & \(7.4347~{\rm GHz}\) \\
Readout-resonator frequency \(\omega_{\rm r}/2\pi\) & \(4.7065~{\rm GHz}\) & \(4.7065~{\rm GHz}\) \\
Readout integration length & \(0.5~\mu{\rm s}\) & \(0.5~\mu{\rm s}\) \\
Local qutrit \(\pi\)-pulse length & \(20~{\rm ns}\) & \(20~{\rm ns}\) \\
Thermal population & \(1.50\pm0.14\%\) & \(1.57\pm0.04\%\) \\
Measured loaded linewidth \(\kappa/2\pi\) & \(10.72~{\rm MHz}\) & \(10.28~{\rm MHz}\) \\
Internal loss rate \(\kappa_{\rm int}/2\pi\) & \(0.0208~{\rm MHz}\) & \(0.0176~{\rm MHz}\) \\
\(T_{1,ge}\) & \(21.78~\mu{\rm s}\) & \(22.16~\mu{\rm s}\) \\
\(T_{1,ef}\) & \(9.79~\mu{\rm s}\) & \(9.75~\mu{\rm s}\) \\
\(T_{\phi,ge}\) & \(3.80~\mu{\rm s}\) & \(2.20~\mu{\rm s}\) \\
\(T_{\phi,ef}\) & \(3.75~\mu{\rm s}\) & \(2.20~\mu{\rm s}\) \\
Channel efficiency \(\eta_c\) & \multicolumn{2}{c}{\(0.92\)} \\
\hline
\end{tabular}
\end{table}

Table~\ref{tab:supp_device_parameters} collects the operating-point parameters used in the main experiment and the cascaded-master-equation simulations. The purpose of the table is to make clear which quantities are independently calibrated before the qutrit-link benchmarks are evaluated. The values of \(T_1\), \(T_\phi\), \(\eta_c\), and \(\kappa_{\rm int}/2\pi\) are the values used in the current simulations.

The thermal populations are the independently calibrated residual excited-state populations at the operating point. The local qutrit control pulses use \(20~{\rm ns}\) calibrated \(\pi\) rotations on both nodes, and the readout integration time is \(0.5~\mu{\rm s}\). The photon-mediated transfer primitives use chirped \(200~{\rm ns}\) sine envelopes.

\subsection{Device coherence, readout calibration, and simulation inputs}
\label{sec:supp_device_coherence_readout_simulation}

\begin{figure}[htbp]
    \centering
    \suppfiginclude{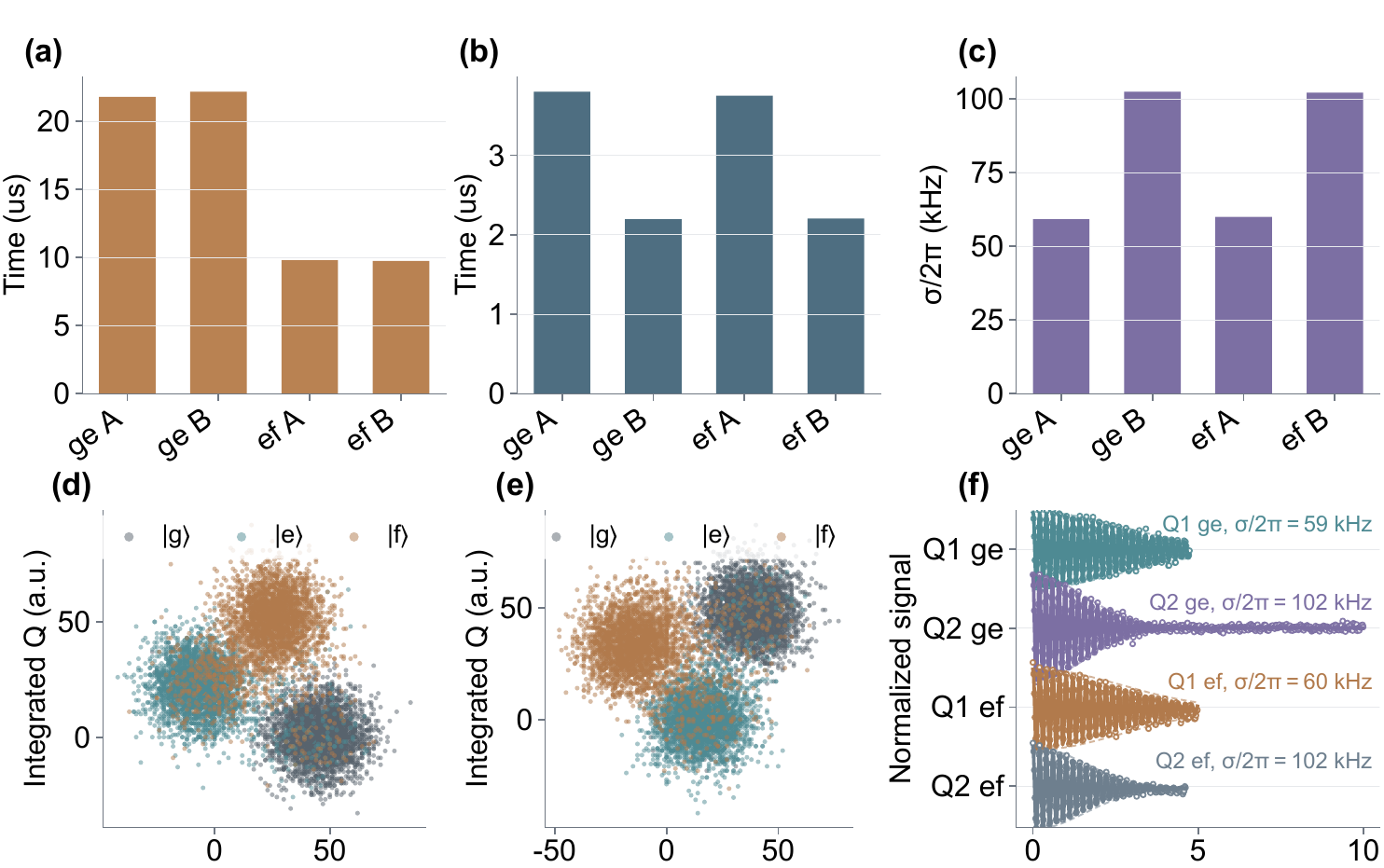}{width=0.95\textwidth}
    \caption{\textbf{Device coherence, qutrit readout calibration, and simulation inputs.}
    \textbf{(a)} Energy-relaxation times \(T_1\) for the \(ge\) and \(ef\) transitions of the two nodes.
    \textbf{(b)} Ramsey-derived pure-dephasing times \(T_\phi\) after subtracting the relaxation-limited contribution.
    \textbf{(c)} Corresponding quasi-static detuning-noise widths \(\sigma/2\pi\) used in the \(1/f\)-like dephasing ensemble.
    \textbf{(d,e)} Representative three-state IQ readout clouds for node A and node B, showing separation of the \(\ket{g}\), \(\ket{e}\), and \(\ket{f}\) states.
    \textbf{(f)} \(T_1\)-corrected Ramsey fits used to extract the quasi-static detuning-noise widths; dashed curves indicate the Gaussian envelope produced by the quasi-static \(1/f\)-like detuning-noise model.}
    \label{fig:supp_device_coherence_readout_simulation}
\end{figure}

Figure~\ref{fig:supp_device_coherence_readout_simulation} collects the device parameters that enter the tomography and simulation. Both nodes use asymmetric-junction transmons, and their operating flux biases are chosen to match the photon interfaces. At these biases, the \(\ket{g}\leftrightarrow\ket{e}\) relaxation times are \(T_{1,ge}=21.78~\mu{\rm s}\) for node A and \(22.16~\mu{\rm s}\) for node B. The \(\ket{e}\leftrightarrow\ket{f}\) transition relaxes faster, with \(T_{1,ef}=9.79~\mu{\rm s}\) and \(9.75~\mu{\rm s}\), respectively. We use these measured values directly as energy-relaxation collapse rates in the cascaded master-equation model.

The Ramsey envelopes decay faster than energy relaxation alone. We attribute the remaining decay to low-frequency, \(1/f\)-like detuning noise and treat it in the quasi-static limit. Each Ramsey trace is fitted with
\begin{equation}
P(t)=y_0+A\exp\left[-\Gamma_{\rm rad}t-\frac{(\sigma t)^2}{2}\right]\cos(2\pi f t+\phi),
\label{eq:supp_ramsey_fit}
\end{equation}
where \(\Gamma_{\rm rad}\) is fixed by the measured \(T_1\) values. For a \(ge\) coherence,
\begin{equation}
\Gamma_{{\rm rad},ge}=\frac{1}{2T_{1,ge}},
\label{eq:supp_gamma_ge}
\end{equation}
whereas for an \(ef\) coherence,
\begin{equation}
\Gamma_{{\rm rad},ef}=\frac{1}{2}\left(\frac{1}{T_{1,ge}}+\frac{1}{T_{1,ef}}\right).
\label{eq:supp_gamma_ef}
\end{equation}
After the \(T_1\) contribution is fixed, the Gaussian part of the envelope defines the rms quasi-static detuning width \(\sigma\). This form gives a better description of the measured Ramsey traces than a single Markovian exponential decay and is the dephasing model used in the main simulations. The extracted widths are \(\sigma_{ge,A}/2\pi=59.24~{\rm kHz}\), \(\sigma_{ge,B}/2\pi=102.47~{\rm kHz}\), \(\sigma_{ef,A}/2\pi=60.01~{\rm kHz}\), and \(\sigma_{ef,B}/2\pi=102.18~{\rm kHz}\), corresponding to \(T_\phi=\sqrt{2}/\sigma\) values of \(3.80~\mu{\rm s}\), \(2.20~\mu{\rm s}\), \(3.75~\mu{\rm s}\), and \(2.20~\mu{\rm s}\).

The IQ-cloud measurements in Fig.~\ref{fig:supp_device_coherence_readout_simulation}(d,e) show that the three qutrit levels are resolved by the heterodyne readout chain. The assignment matrices used to correct tomography probabilities are shown separately in Fig.~\ref{fig:supp_readout_correction}, where the full two-qutrit correction matrices are also displayed.

\subsection{Cavity spectroscopy and tunable matching of the two nodes}
\label{sec:supp_cavity_spectroscopy_matching}

\begin{figure}[htbp]
    \centering
    \suppfiginclude{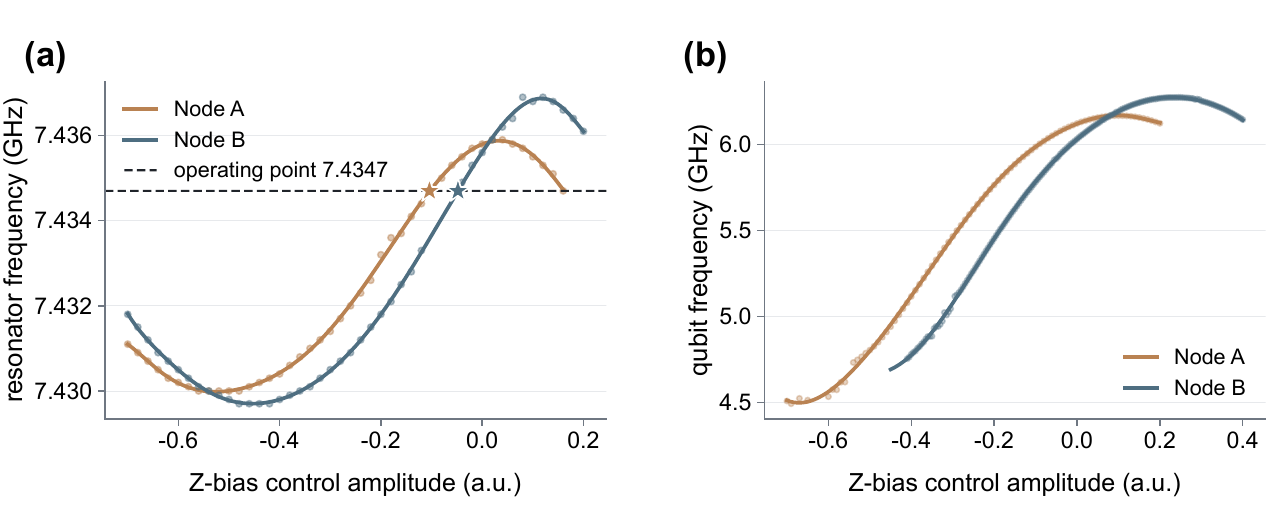}{width=0.95\textwidth}
    \caption{\textbf{Cavity spectroscopy and tunable matching of the two remote interfaces.}
    \textbf{(a)} Transmission-resonator frequency trajectories as a function of the \(Z\)-bias control amplitude. The dashed line marks the matched operating frequency near \(7.4347~{\rm GHz}\), and stars mark the negative-bias operating points used in the experiment.
    \textbf{(b)} Qubit spectroscopy trajectories extracted from the corresponding transmon spectroscopy maps, showing the large flux-tuning range of the asymmetric-junction transmons. Both horizontal axes show the \(Z\)-bias control amplitude in arbitrary units.}
    \label{fig:supp_cavity_spectroscopy_matching}
\end{figure}

The cavity spectroscopy in Fig.~\ref{fig:supp_cavity_spectroscopy_matching} is used to choose the two remote interface frequencies. Each auxiliary resonator is dispersively shifted by its transmon, whose asymmetric SQUID loop provides a tuning range on the order of \(2~{\rm GHz}\). The qutrit flux bias therefore moves the effective auxiliary-resonator frequency, while the Purcell-filter bias controls the coupling rate to the waveguide. These two controls are calibrated separately before the state-transfer experiment.

In the cascaded transfer picture, the photon must be matched twice: its carrier frequency must be resonant with the receiver interface, and its temporal envelope must match the receiver's time-dependent absorption mode. Frequency matching fixes the carrier; linewidth matching fixes the bandwidth scale of the emitted and absorbed wave packets. Both are needed before the pulse-shape calibration can be meaningful.

The final operating frequency is determined from matched spectroscopy line cuts and lies near \(7.4347~{\rm GHz}\). The two stars in Fig.~\ref{fig:supp_cavity_spectroscopy_matching}(a) mark the operating points used in the experiment.

The qubit spectroscopy in Fig.~\ref{fig:supp_cavity_spectroscopy_matching}(b) is extracted by taking the maximum spectroscopy response in each flux-bias row of the auto-asymmetry scans. It shows that both asymmetric-junction transmons can be tuned over a range of about \(2~{\rm GHz}\), large enough to bring the two dressed transmission-resonator frequencies into resonance while keeping the qutrit operating points away from the same nominal flux bias. The corresponding linewidth tunability and matched high-bandwidth operating points are shown in the following section.

\subsection{Tunable Purcell-filter control of the transmission-resonator linewidth}
\label{sec:supp_tunable_purcell_kappa}

Each tunable Purcell filter is a half-wavelength resonator with a SQUID inserted at a current antinode. Magnetic flux applied through the dedicated filter \(z\)-control line changes the SQUID's effective Josephson inductance and thereby tunes the filter-resonance frequency. This SQUID-based frequency-tuning approach follows established tunable-resonator designs~\cite{supp:wallquist2006selective,supp:sandberg2008tuning,supp:wang2013quantum}. Changing the detuning between the filter and the transmission resonator controls the filter-mediated radiative coupling to the microwave link and hence the transmission resonator's effective decay rate. The filter bias is therefore the bandwidth-control knob, while the qutrit bias is used for carrier-frequency matching.

\begin{figure}[htbp]
    \centering
    \suppfiginclude{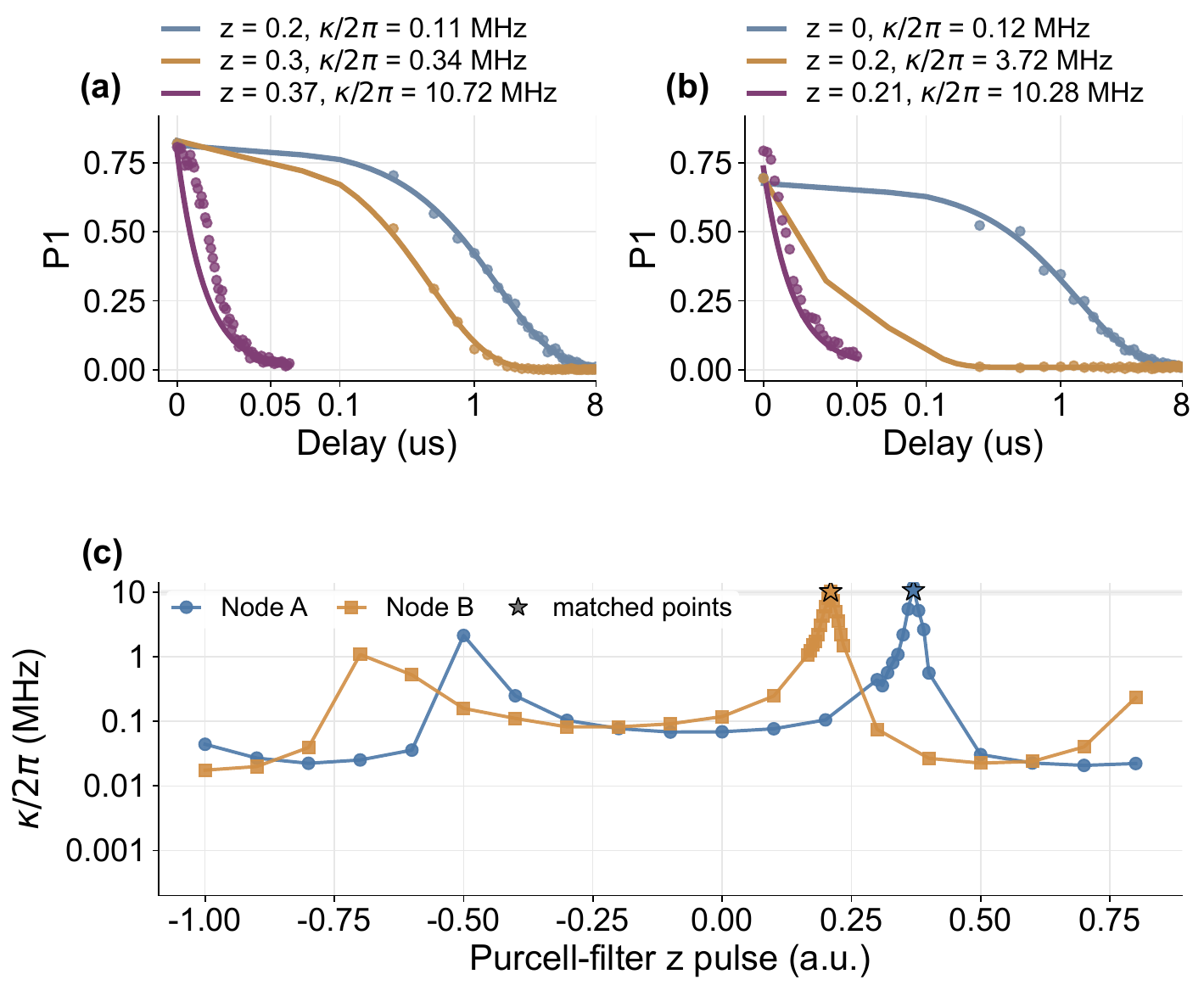}{width=0.95\textwidth}
    \caption{\textbf{Tunable Purcell-filter control of the photon bandwidth.}
    \textbf{(a,b)} Representative cavity-energy decay measurements for node A and node B at different Purcell-filter control amplitudes. Solid curves are single-exponential fits.
    \textbf{(c)} Extracted tunable coupling decay rate \(\kappa/2\pi=1/(2\pi T_{\rm cav})\) as a function of the Purcell-filter \(z\)-pulse amplitude. Stars mark the nearest directly measured high-bandwidth settings used to tune the two remote interfaces to comparable photon bandwidths.}
    \label{fig:supp_tunable_purcell_kappa}
\end{figure}

In addition to frequency matching, the experiment uses tunable Purcell filters to control the coupling between each auxiliary resonator and the common microwave link~\cite{supp:houck2008controlling,supp:reed2010fast,supp:bronn2015broadband,supp:yan2023broadband}. We calibrate this control by preparing one photon in the auxiliary resonator and measuring its energy decay as a function of the Purcell-filter control amplitude. For each trace, we fit the measured population to \(P_1(t)=P_{\rm off}+A\exp(-t/T_{\rm cav})\), and define the loaded photon linewidth as \(\kappa/2\pi=1/(2\pi T_{\rm cav})\).

The Purcell-filter bias tunes the loaded transmission-resonator linewidth from the \(10^{-2}\)-MHz range to above \(10~{\rm MHz}\), as shown in Fig.~\ref{fig:supp_tunable_purcell_kappa}. The nearest directly measured settings to \(10~{\rm MHz}\) are \(\kappa_A/2\pi=10.72\pm0.44~{\rm MHz}\) at \(z=0.37\) and \(\kappa_B/2\pi=10.28\pm0.26~{\rm MHz}\) at \(z=0.21\). Over the measured range the linewidth changes from \(0.0208~{\rm MHz}\) to \(12.16~{\rm MHz}\) for node A and from \(0.0176~{\rm MHz}\) to \(10.28~{\rm MHz}\) for node B, corresponding to an on/off ratio of about \(5.8\times10^2\).

The high-bandwidth points calibrate the interface response under the applied control pulses. The decay measurements use nominal square Purcell-filter \(z\)-pulses. At the largest coupling settings, the cavity lifetime is only \(\sim15~{\rm ns}\), comparable to the rise time, filtering and settling of the control line. The fitted decay therefore includes the response of the Purcell-filter pulse itself. This explains why the points near \(z=0.37\) for node A and \(z=0.21\) for node B do not follow a simple exponential dependence on the programmed amplitude. For the transfer experiment, the relevant outcome is that both interfaces can be biased to comparable photon bandwidths after their carrier frequencies have been matched.

The smallest loaded linewidths in the same calibration provide an estimate of the intrinsic auxiliary-resonator loss used in the simulations. From the longest measured cavity lifetimes, \(T_{{\rm cav},A}=7.645~\mu{\rm s}\) and \(T_{{\rm cav},B}=9.057~\mu{\rm s}\), we obtain \(\kappa_{{\rm int},A}/2\pi=0.0208~{\rm MHz}\) and \(\kappa_{{\rm int},B}/2\pi=0.0176~{\rm MHz}\), corresponding to collapse rates \(0.1308~\mu{\rm s}^{-1}\) and \(0.1104~\mu{\rm s}^{-1}\), respectively. These values are used as fixed internal-loss collapse rates in all pulse-level simulations.

\subsection{Chirped-pulse calibration for the transfer primitives}
\label{sec:supp_chirped_pulse_calibration}

\begin{figure}[htbp]
    \centering
    \suppfiginclude{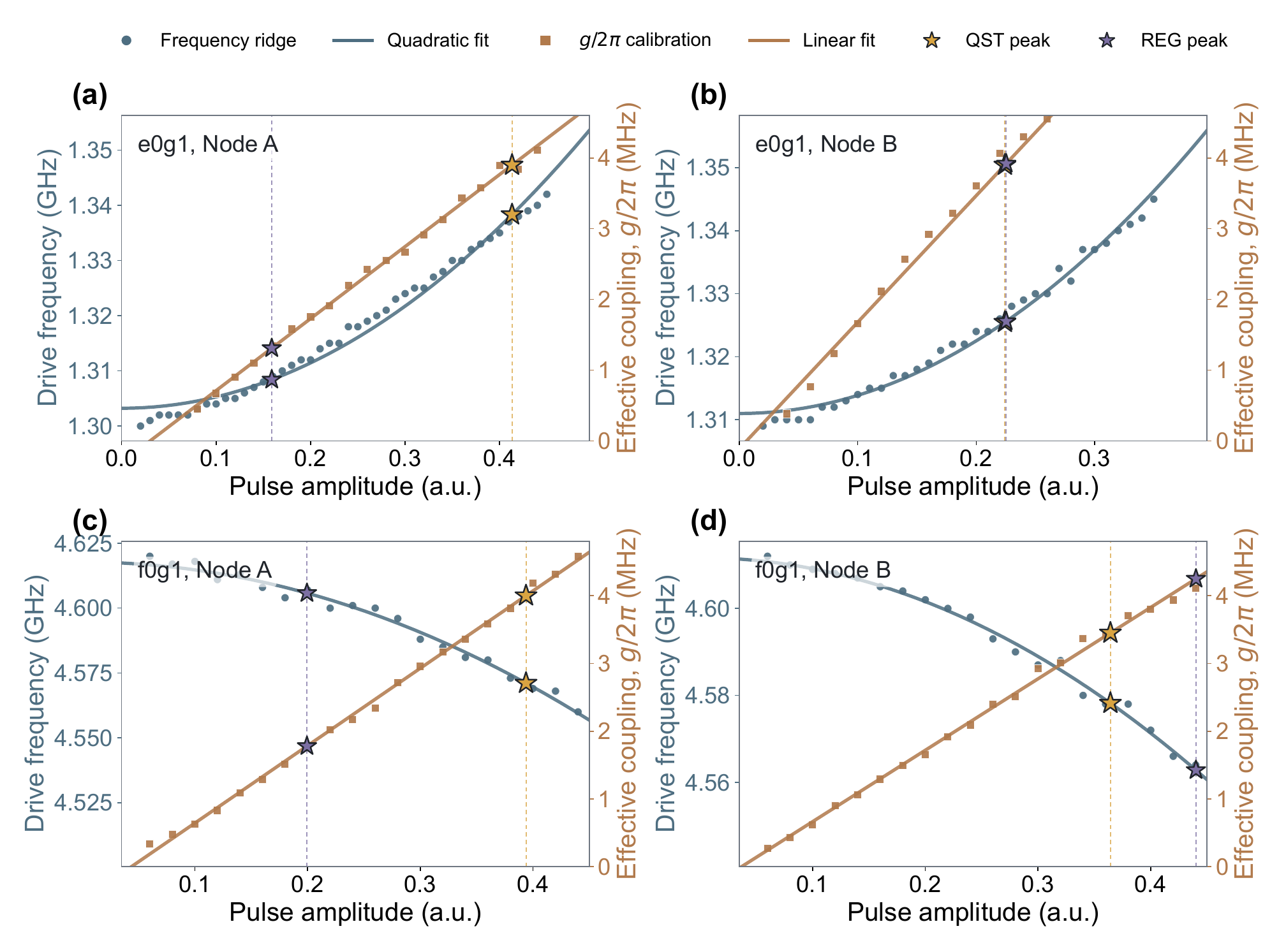}{width=0.95\textwidth}
    \caption{\textbf{Amplitude calibration for the chirped \(e0g1\) and \(f0g1\) primitives.}
    \textbf{(a,b)} \(e0g1\) calibrations for nodes A and B.
    \textbf{(c,d)} \(f0g1\) calibrations for nodes A and B.
    Blue circles, read on the left axes, show the extracted resonance frequency from the amplitude--frequency spectroscopy scans; solid blue curves are quadratic fits. Brown squares, read on the right axes, show the independently measured effective coupling \(g/2\pi\) extracted from the first \(P_{\rm initial}=0.5\) crossing of the Rabi traces; solid brown curves are linear fits. Gold and purple stars mark the programmed pulse amplitudes used in the QST and REG experiments. At those amplitudes, the stars on the brown calibration curves give the corresponding effective couplings, while the stars on the blue quadratic fits give the chirped drive frequencies. The dashed vertical guides connect the two calibrations.}
    \label{fig:supp_chirped_pulse_calibration}
\end{figure}

The transfer pulses used in the experiment have a fixed \(200~{\rm ns}\) sine amplitude envelope, but their carrier frequencies are not held fixed during the pulse. The reason is the same for both transfer primitives: the interaction that turns on the photon-mediated coupling also shifts the instantaneous resonance condition. For the \(e0g1\) primitive, the flux-parametric modulation activates the \(\ket{e,0}\leftrightarrow\ket{g,1}\) sideband, but the finite modulation amplitude also changes the time-averaged transmon frequency. In the language used in our earlier shaped-photon experiment~\cite{supp:li2025ondemandshaped}, this produces an amplitude-dependent sideband detuning through the combined dc frequency shift and the small sideband-induced Lamb shift. For the \(f0g1\) primitive, the cavity-assisted Raman drive produces an analogous amplitude-dependent ac Stark shift of the Raman resonance~\cite{supp:magnard2018fast,supp:kurpiers2018deterministic,supp:magnard2020microwave}. If this detuning were left uncompensated, the beginning, middle and end of the shaped pulse would not be resonant at the same drive frequency. The resulting phase accumulation would distort the emitted photon and reduce the capture efficiency.

We therefore calibrate a dynamic frequency correction for each node and each primitive. For a given primitive \(p\in\{e0g1,f0g1\}\), we apply a flat-top calibration pulse with amplitude \(A\), sweep the modulation or Raman-drive frequency, and record the population transfer after the pulse. The resonance appears as a ridge in the two-dimensional maps in Fig.~\ref{fig:supp_chirped_pulse_calibration}. The lowest-amplitude point is excluded from the fit because the transfer contrast is weak and the ridge position is more sensitive to background subtraction. The remaining ridge is fit to
\begin{equation}
\omega_{p,j}(A)=\omega_{p,j}^{(0)}+c_{p,j}A^2 ,
\label{eq:supp_chirp_quadratic}
\end{equation}
where \(j=A,B\) labels the node. We use a quadratic fit without a linear term because both the average flux-induced frequency shift and the leading ac Stark shift are even in the pulse amplitude. This is also the calibration form used in our previous \(e0g1\) shaped-photon work.

During the transfer experiment, the sine envelope \(A(t)\) is converted into a time-dependent drive frequency by evaluating the calibrated ridge frequency at the instantaneous pulse amplitude,
\begin{equation}
\omega_d(t)=\omega_{\rm ridge}\!\left[A(t)\right],
\label{eq:supp_chirped_pulse_frequency}
\end{equation}
or, equivalently, by programming a time-dependent phase
\begin{equation}
\theta_d(t)=\theta_d(0)+\int_0^t \omega_d(t')\,dt' .
\label{eq:supp_chirped_phase}
\end{equation}
In the rotating frame of the unchirped drive, this procedure cancels the amplitude-dependent residual detuning \(\Delta_p(t)=\omega_{p,j}[A(t)]-\omega_{p,j}^{(0)}\) by adding the opposite time-dependent phase accumulation to the control waveform. The pulse is therefore resonant throughout the full emission or capture window. In the pulse-level simulation, the transfer coupling \(g_{e,j}(t)\) or \(g_{f,j}(t)\) is defined in this chirp-corrected resonant frame, while residual detuning noise is included separately through the quasi-static dephasing model described above.

This frequency-ridge calibration is distinct from the amplitude-to-coupling calibration shown on the right axes of Fig.~\ref{fig:supp_chirped_pulse_calibration}. For a fixed chirp-corrected resonance condition, the \(e0g1\) flux-modulation amplitude and the \(f0g1\) Raman-drive amplitude set the corresponding resonant sideband rates. In the small-amplitude range used for the partial-transfer pulses, the measured transfer rate is linear in the programmed drive amplitude, so the effective peak couplings \(g_{e,j}^{\rm max}\) and \(g_{f,j}^{\rm max}\) used in the simulations are obtained by scaling the calibrated full-transfer couplings. The higher-amplitude full-transfer pulses are calibrated directly from their transfer oscillations, while the chirp calibration above keeps each instantaneous amplitude resonant during the shaped pulse.

\section{\MakeUppercase{Transfer primitives and tomography reconstruction}}
\label{sec:supp_transfer_readout_tomography}
\label{sec:supp_tomography_readout_robustness}

The benchmarks in the main text are obtained from qutrit and two-qutrit tomography. This section therefore separates three logically different ingredients: timing and phase calibrations for the transfer primitives, the fixed readout-assignment correction applied to measured probabilities, and the reconstructed matrices used as diagnostics of the primitive operations. The readout correction uses a linear inversion of independently measured assignment matrices, followed by the same tomography reconstruction for every input state and every repetition.

\subsection{Primitive delay and phase calibrations}
\label{sec:supp_primitive_transfer_alternate}

\begin{figure}[htbp]
    \centering
    \suppfiginclude{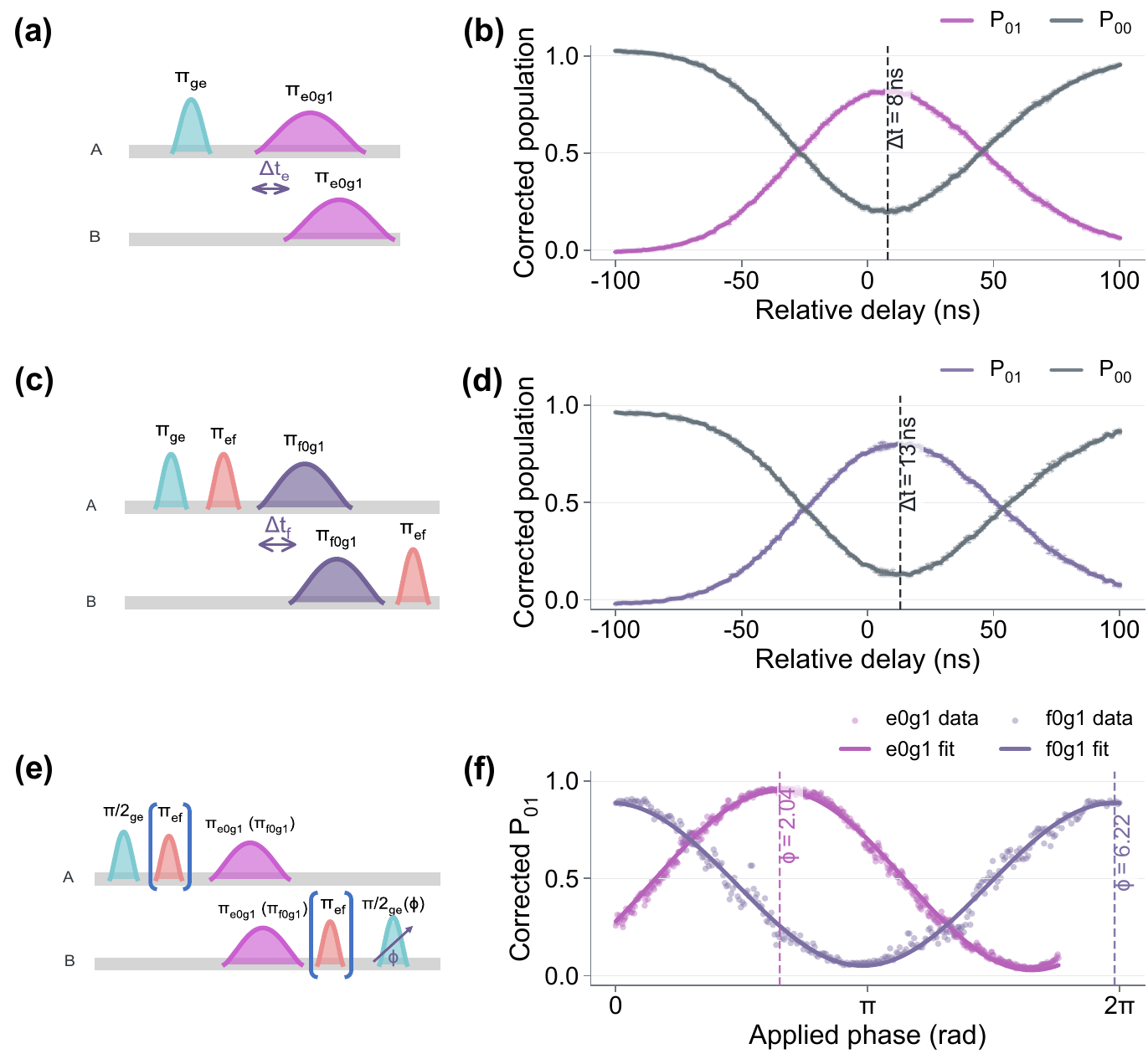}{width=0.95\textwidth}
    \caption{\textbf{Primitive delay and phase calibrations for \(e0g1\) and \(f0g1\) transfer.}
    \textbf{(a,c,e)} Pulse sequences for the \(e0g1\) delay, \(f0g1\) delay and Ramsey-type phase calibrations.
    \textbf{(b,d)} Readout-corrected relative-delay scans for \(e0g1\) and \(f0g1\), respectively.
    \textbf{(f)} Ramsey phase scan used to determine the applied phase settings.}
    \label{fig:supp_primitive_transfer_alternate}
\end{figure}

The delay and phase scans in Fig.~\ref{fig:supp_primitive_transfer_alternate} set the timing and phase frame used in the qutrit QPT and REG sequences. For the \(e0g1\) delay calibration, node A is prepared by a local \(\pi_{ge}\) pulse, emits through a \(\pi_{e0g1}\) pulse, and node B receives with a delayed \(\pi_{e0g1}\) pulse. For the \(f0g1\) delay calibration, node A is prepared by \(\pi_{ge}\) and \(\pi_{ef}\) pulses, emits through a \(\pi_{f0g1}\) pulse, and node B receives with a delayed \(\pi_{f0g1}\) pulse followed by \(\pi_{ef}\) mapping. The Ramsey-type phase calibration scans the receiver analysis pulse \(\pi/2_{ge}(\phi)\), compensating the deterministic dynamical phase accumulated during photon propagation and capture. The selected cutoff-time dynamics are shown in the main text, so here we show the calibration scans instead. The Ramsey phase scan in Fig.~\ref{fig:supp_primitive_transfer_alternate}(f) records the readout-corrected receiver population \(P_{01}\) as a function of applied phase; the fitted sinusoid defines the virtual-\(Z\) reference.

For \(e0g1\) transfer, the experiment starts from an excitation in node A and monitors the sender population \(P_{10}\), receiver population \(P_{01}\), and ground-state population \(P_{00}\) as the relative timing between the two shaped pulses is varied. The optimal delay is chosen near the maximum of the transferred population and corresponds to the temporal overlap used in the main-text dynamics. For \(f0g1\) transfer, the initial excitation is in the second excited state of node A. A final local \(\pi_{ef}\) pulse maps the receiver-side \(f\)-state population into the \(e\)-state readout manifold, following the experimental sequence.

The solid simulation curves in the corresponding main-text figure are obtained from the same pulse-level cascaded master equation used for the QST, QPT, and REG simulations. The \(e0g1\) and \(f0g1\) primitive transfer curves are simulated using \(200~{\rm ns}\) sine pulses, \(\eta_c=0.92\), \(\kappa_{{\rm int},A}/2\pi=0.0208~{\rm MHz}\), \(\kappa_{{\rm int},B}/2\pi=0.0176~{\rm MHz}\), measured \(T_1\) values, and the \(T_1\)-corrected quasi-static dephasing ensemble. No population trace is shifted separately in this comparison.

\subsection{Logical state maps for qutrit transfer and REG}
\label{sec:supp_protocol_state_maps}

We spell out the ideal logical maps implemented by the qutrit-transfer and REG pulse sequences. These maps omit deterministic phases from photon propagation, shaped emission and capture, and local qutrit rotations. In the experiment, these phases are absorbed into calibrated drive phases and virtual-\(Z\) corrections before tomography.

For qutrit-state transfer, we write \(\ket{n_1 n_2}\) for the photon-number state of the first and second temporal modes. The first temporal mode is generated and absorbed by the \(f0g1\) primitive, while the second temporal mode is generated and absorbed by the \(e0g1\) primitive. The emission stage maps
\begin{equation}
\begin{aligned}
(\alpha\ket{g}+\beta\ket{e}+\gamma\ket{f})_A\ket{g}_B\ket{0_1 0_2}
&\xrightarrow{\pi_{f0g1}}
\left(\alpha\ket{g}_A+\beta\ket{e}_A\right)\ket{g}_B\ket{0_1 0_2}
+\gamma\ket{g}_A\ket{g}_B\ket{1_1 0_2}\\
&\xrightarrow{\pi_{e0g1}}
\ket{g}_A\ket{g}_B
\left(\alpha\ket{0_1 0_2}
+\gamma\ket{1_1 0_2}
+\beta\ket{0_1 1_2}\right).
\end{aligned}
\label{eq:supp_qutrit_transfer_time_bin_map}
\end{equation}
The receiver applies the time-reversed capture operations to the same temporal modes,
\begin{equation}
\ket{g}_A\ket{g}_B
\left(\alpha\ket{0_1 0_2}
+\gamma\ket{1_1 0_2}
+\beta\ket{0_1 1_2}\right)
\rightarrow
\ket{g}_A(\alpha\ket{g}+\beta\ket{e}+\gamma\ket{f})_B .
\label{eq:supp_qutrit_transfer_capture_map}
\end{equation}
The qutrit is therefore carried by a hybrid encoding: \(\ket{g}\) is the two-bin vacuum, \(\ket{f}\) is a single photon in the first bin, and \(\ket{e}\) is a single photon in the second bin.

For REG, the sender pulses are partial transfers. With the convention that a pulse angle \(\theta\) produces amplitudes \(\cos(\theta/2)\) and \(\sin(\theta/2)\), the ideal sequence starts from \(\ket{e}_A\ket{g}_B\ket{0_1 0_2}\). The first partial \(e0g1\) pulse has \(\theta_e=2\arctan(1/\sqrt{2})\), so that
\begin{equation}
\ket{e}_A\ket{g}_B\ket{0_1 0_2}
\rightarrow
\sqrt{\frac{2}{3}}\ket{e}_A\ket{g}_B\ket{0_1 0_2}
+\frac{1}{\sqrt{3}}\ket{g}_A\ket{g}_B\ket{1_1 0_2}.
\label{eq:supp_reg_first_partial}
\end{equation}
Local \(\pi_{ef}\) and \(\pi_{ge}\) rotations on node A relabel the stored branches,
\begin{equation}
\sqrt{\frac{2}{3}}\ket{f}_A\ket{g}_B\ket{0_1 0_2}
+\frac{1}{\sqrt{3}}\ket{e}_A\ket{g}_B\ket{1_1 0_2}.
\label{eq:supp_reg_local_relabel}
\end{equation}
A subsequent \(f0g1\) partial pulse with \(\theta_f=\pi/2\) gives
\begin{equation}
\frac{1}{\sqrt{3}}\ket{f}_A\ket{g}_B\ket{0_1 0_2}
+\frac{1}{\sqrt{3}}\ket{g}_A\ket{g}_B\ket{0_1 1_2}
+\frac{1}{\sqrt{3}}\ket{e}_A\ket{g}_B\ket{1_1 0_2}.
\label{eq:supp_reg_three_path_flying_state}
\end{equation}
Full capture of the first bin by \(e0g1\) and of the second bin by \(f0g1\) maps the three branches to
\begin{equation}
\ket{\Psi_{\rm qutrit}}=
\frac{1}{\sqrt{3}}\left(\ket{fg}+\ket{gf}+\ket{ee}\right),
\label{eq:supp_reg_capture_map}
\end{equation}
up to local phases. This protocol contains no \(\ket{1_1 1_2}\) flying-photon component. The entanglement is generated through three coherent alternatives: no photon, a first-bin photon, or a second-bin photon. Therefore at most one photon is exposed to propagation loss in the link.

\subsection{Tomography measurement bases and readout correction}
\label{sec:supp_readout_correction}

\begin{figure}[htbp]
    \centering
    \suppfiginclude{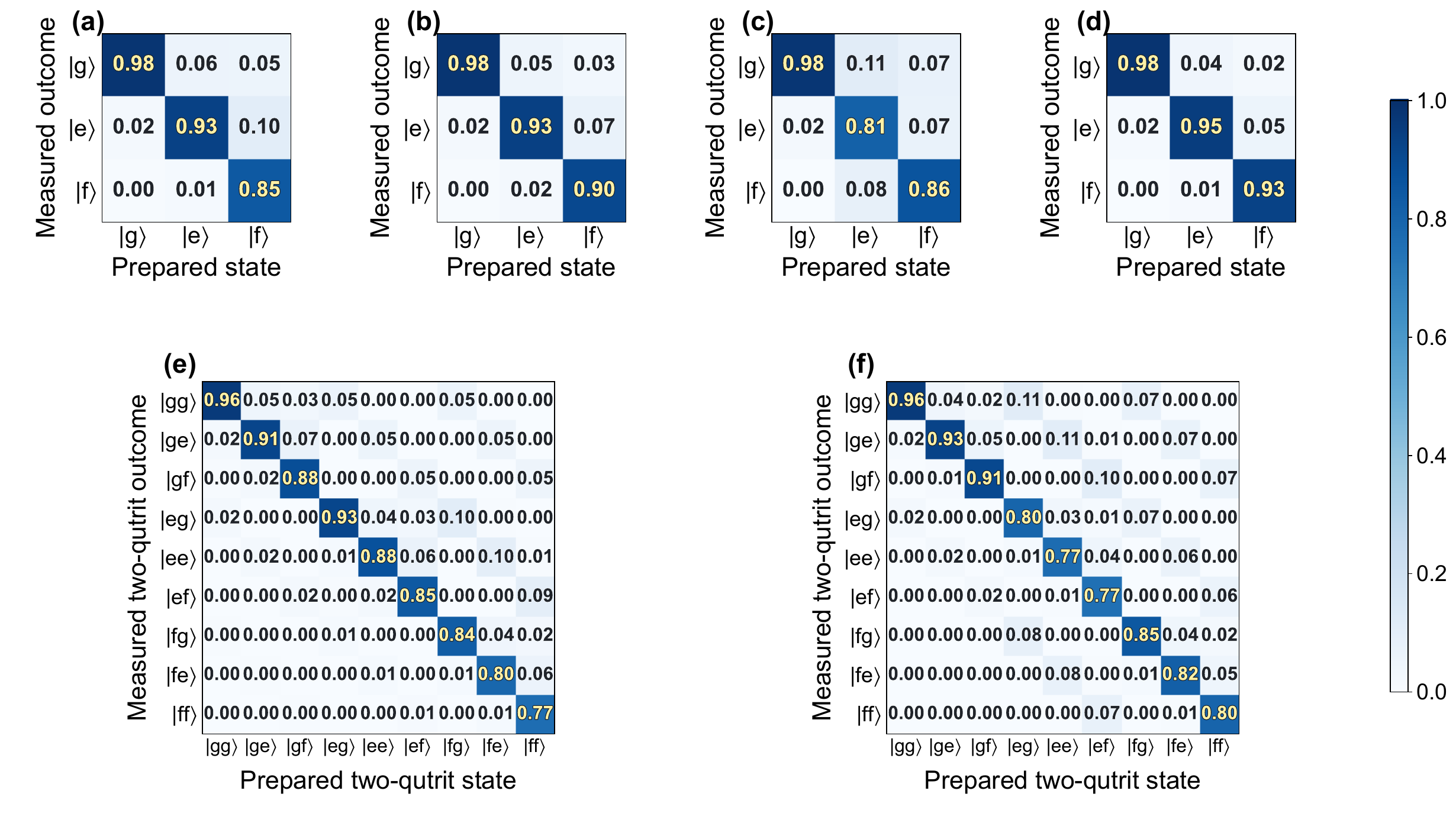}{width=0.95\textwidth}
    \caption{\textbf{Qutrit readout correction.}
    \textbf{(a,b)} Single-qutrit readout assignment matrices measured before the qutrit state-transfer QST/QPT experiment and used for that reconstruction.
    \textbf{(c,d)} Single-qutrit readout assignment matrices measured before the REG tomography experiment and used for that reconstruction.
    \textbf{(e,f)} Full two-qutrit correction matrices \(M_A\otimes M_B\) used for the corresponding tomography datasets. Numerical values are written in each matrix element to make the correction transparent.}
    \label{fig:supp_readout_correction}
\end{figure}

Figure~\ref{fig:supp_readout_correction} gives the assignment matrices used for heterodyne qutrit readout and state discrimination~\cite{supp:Silva2010schemes,supp:caves1982quantum}. These matrices are independent calibration data. Panels (a) and (b) were measured before the qutrit state-transfer QST/QPT datasets, and panels (c) and (d) were measured before the REG tomography dataset, using the same readout pulses, integration window and state-discrimination procedure as the corresponding tomography run. Once measured, the assignment matrices are kept fixed for every input state, every analysis basis and every repetition in that run. No entry of the matrices is fitted to the reconstructed density matrix, to the ideal target state or to the beyond-qubit benchmarks.

Three-level readout errors are the dominant SPAM imperfection in the tomography data, so the assignment matrices are measured in the same experimental configuration as the tomography experiments and then applied as a fixed linear response correction. Before readout, tomography pulses rotate the desired qutrit coherence into a population difference. For one qutrit we use the nine analysis settings
\begin{equation}
\begin{aligned}
\mathcal{A}_{1{\rm q}}=\{&
I,\ R^{ge}_{x}(\pi/2),\ R^{ge}_{y}(\pi/2),\ R^{ef}_{x}(\pi/2),\ R^{ef}_{y}(\pi/2),\\
&R^{ef}_{x}(\pi/2)R^{ge}_{x}(\pi/2),\
R^{ef}_{y}(\pi/2)R^{ge}_{x}(\pi/2),\
R^{ef}_{x}(\pi/2)R^{ge}_{y}(\pi/2),\
R^{ef}_{y}(\pi/2)R^{ge}_{y}(\pi/2)\}.
\end{aligned}
\label{eq:supp_qutrit_tomo_basis}
\end{equation}
Here \(R^{mn}_{x,y}(\pi/2)\) denotes a calibrated \(\pi/2\) rotation in the \(\{\ket{m},\ket{n}\}\) subspace. This overcomplete analysis set accesses the populations and the real and imaginary parts of \(\rho_{ge}\), \(\rho_{gf}\) and \(\rho_{ef}\). The non-adjacent \(\ket{g}\)-\(\ket{f}\) coherence is not measured with a direct \(gf\) pulse; it is mapped to a population difference through composite rotations on the adjacent \(ge\) and \(ef\) transitions. After readout correction, the density matrix is obtained by maximum-likelihood estimation constrained to be Hermitian, positive semidefinite and unit trace.

For two-qutrit state tomography, the single-qutrit analysis settings are applied independently to both nodes. This gives \(9^2=81\) joint analysis settings. For each setting, the two qutrits are read out in the product basis \(\{\ket{gg},\ket{ge},\ket{gf},\ket{eg},\ldots,\ket{ff}\}\), producing a nine-component probability vector. These probability vectors reconstruct the full \(9\times9\) density matrix used for the REG fidelity, negativity~\cite{supp:peres1996separability,supp:vidal2002computable,supp:horodecki2009quantum}, dense-coding capacity~\cite{supp:bennett1992communication,supp:horodecki2001classical,supp:bowen2001classical} and CGLMP calculations~\cite{supp:collins2002bell}.

The measured qutrit population vector \(\mathbf{p}_{\rm meas}\) is related to the corrected population vector \(\mathbf{p}\) by
\begin{equation}
\mathbf{p}_{\rm meas}=M\mathbf{p},
\label{eq:supp_assignment_matrix}
\end{equation}
where \(M_{ij}\) is the probability to assign the measurement outcome \(i\) when the prepared state is \(j\). The corrected probabilities are obtained as
\begin{equation}
\mathbf{p}=M^{-1}\mathbf{p}_{\rm meas}.
\label{eq:supp_readout_inverse}
\end{equation}
For two-qutrit tomography, the assignment matrix is
\begin{equation}
M_{AB}=M_A\otimes M_B .
\label{eq:supp_two_qutrit_readout}
\end{equation}
The same correction procedure is used for qutrit state-transfer QPT and for two-qutrit REG tomography. In qutrit process tomography, nine linearly independent qutrit input states are prepared on node A using calibrated local rotations in the \(\ket{g}\)-\(\ket{e}\) and \(\ket{e}\)-\(\ket{f}\) subspaces. In the data shown in the main text, these states are generated from \(\ket{g}\) by the gate pairs
\begin{equation}
\begin{gathered}
(I,I),\ (X_{ge},I),\ (X_{ge},X_{ef}),\
(X_{ge}/2,I),\ (X_{ge}/2,X_{ef}),\\
(X_{ge},X_{ef}/2),\ (Y_{ge}/2,I),\
(Y_{ge}/2,X_{ef}),\ (X_{ge},Y_{ef}/2),
\end{gathered}
\label{eq:supp_qpt_input_states}
\end{equation}
where the first and second entries denote the \(ge\)- and \(ef\)-subspace preparation rotations, respectively. For each input state, we reconstruct the prepared state before transfer and the received state after transfer using the same readout-corrected maximum-likelihood tomography procedure. The resulting input-output density-matrix pairs are used to reconstruct the qutrit process matrix in the Gell-Mann basis. These maximum-likelihood density matrices are used for fidelity and information-metric evaluation.

Figure~\ref{fig:supp_reg_information_metrics} summarizes readout-correction sensitivity and assignment-matrix uncertainty.

\subsection{Primitive transfer and REG QST diagnostics}
\label{sec:supp_primitive_transfer_reg_qst}

\begin{figure}[htbp]
    \centering
    \suppfiginclude{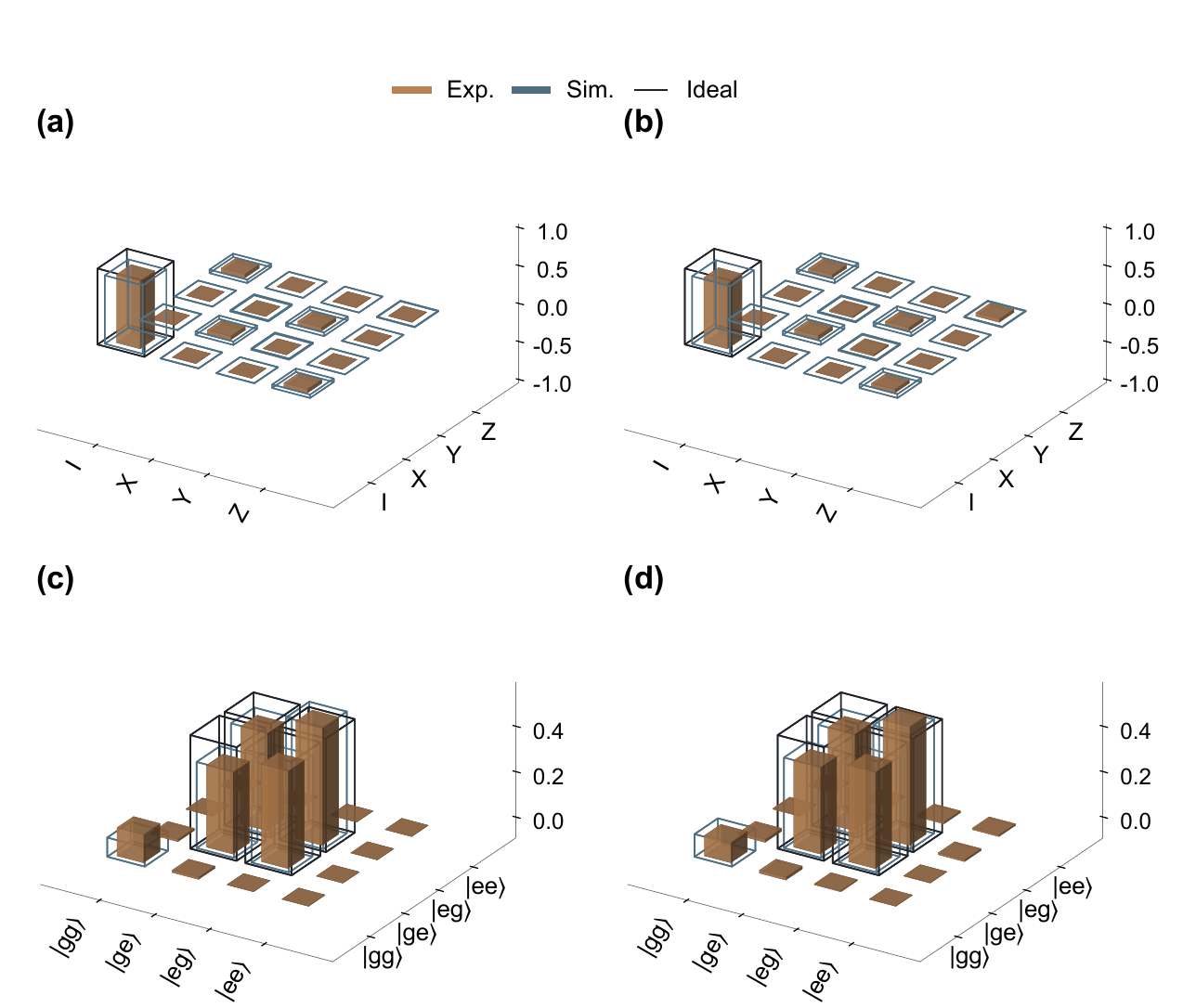}{width=0.95\textwidth}
    \caption{\textbf{Primitive transfer and REG QST diagnostics.}
    \textbf{(a,b)} Effective two-level state-transfer tomography for the \(e0g1\) and \(f0g1\) primitives. Filled bars denote experiment, blue outlines denote pulse-level simulation, and black outlines denote the ideal transfer operation in the corresponding two-level subspace.
    \textbf{(c,d)} Two-qutrit QST of REG states generated with the \(e0g1\) and \(f0g1\) primitives, respectively. These panels show the reconstructed REG density matrices.}
    \label{fig:supp_primitive_transfer_reg_qst}
\end{figure}

Figure~\ref{fig:supp_primitive_transfer_reg_qst} gives tomography diagnostics for the elementary \(e0g1\) and \(f0g1\) operations. Panels (a) and (b) are effective two-level state-transfer reconstructions. The \(e0g1\) primitive transfers the \(\{\ket{g},\ket{e}\}\) subspace. The \(f0g1\) primitive transfers the \(\{\ket{g},\ket{f}\}\) excitation and then maps the receiver population back to the \(\{\ket{g},\ket{e}\}\) readout subspace with a local \(\pi_{ef}\) pulse.

Panels (c) and (d) are two-qutrit QST results for REG sequences constructed from the same primitives. They reconstruct density matrices of partially transferred states. The simulations use the same cascaded master-equation model and calibrated channel parameters as the main-text state-transfer and REG simulations.

These primitive fidelities are diagnostic benchmarks for the building blocks. The high-dimensional channel benchmark is the full qutrit QPT data in the main text and in Fig.~\ref{fig:supp_qutrit_process_tomography}. The primitive diagnostics are included here to show that the same calibrated transfer operations used in the main sequence behave consistently at the level of individual transitions.

\subsection{Phase and matrix-distance diagnostics}
\label{sec:supp_tomography_phase_diagnostics}

\begin{figure}[htbp]
    \centering
    \suppfiginclude{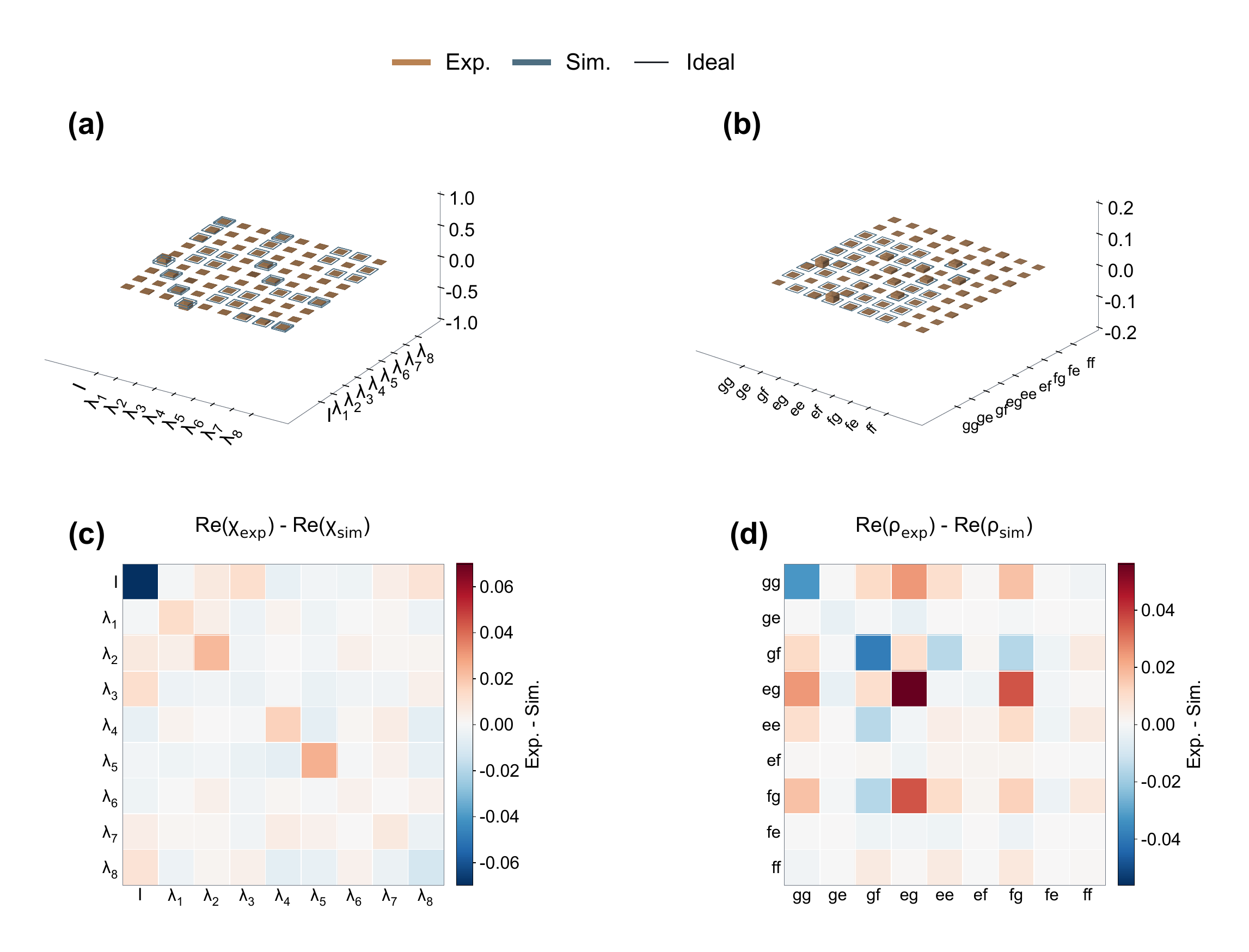}{width=0.98\textwidth,height=0.76\textheight,keepaspectratio}
    \caption{\textbf{Phase and residual matrix diagnostics for qutrit transfer and REG.}
    \textbf{(a)} Imaginary part of the qutrit state-transfer process matrix in the Gell-Mann operator basis.
    \textbf{(b)} Imaginary part of the phase-aligned two-qutrit REG density matrix.
    \textbf{(c)} Residual real part of the qutrit-transfer process matrix, \({\rm Re}(\chi_{\rm exp})-{\rm Re}(\chi_{\rm sim})\).
    \textbf{(d)} Residual real part of the REG density matrix, \({\rm Re}(\rho_{\rm exp})-{\rm Re}(\rho_{\rm sim})\).
    Experimental filled bars, simulated blue outlines and ideal black outlines are overlaid in \textbf{(a,b)}; \textbf{(c,d)} use centered color scales for experiment--simulation residuals.}
    \label{fig:supp_tomography_phase_diagnostics}
    \label{fig:supp_qutrit_process_tomography}
\end{figure}

The real parts of the qutrit process matrix and the REG density matrix carry the main physical signal and are shown in the main text as compact overlaid matrix panels. Figure~\ref{fig:supp_tomography_phase_diagnostics} and Table~\ref{tab:supp_exp_sim_matrix_distance} provide the complementary diagnostics needed to read those panels quantitatively. The imaginary components in Fig.~\ref{fig:supp_tomography_phase_diagnostics}(a,b) are sensitive to residual phase errors after calibration of the transfer frame and local virtual-\(Z\) corrections. The residual maps in Fig.~\ref{fig:supp_tomography_phase_diagnostics}(c,d) show the element-wise real-part differences between the experimental matrices and the corresponding pulse-level simulations, while Table~\ref{tab:supp_exp_sim_matrix_distance} reports the associated matrix distances.

The qutrit process matrix is reconstructed in the Gell-Mann operator basis from the measured input-output state pairs. The experimental process fidelity is \(77.12\%\pm0.10\%\), and the average final-state fidelity over the nine tested qutrit states is \(83.68\%\). The simulated values are \(84.10\%\) and \(89.45\%\), respectively. Both experiment and simulation exceed the qubit-channel bound for average qutrit state transfer, i.e. the maximum Haar-averaged fidelity \(3/4\) obtainable when an arbitrary qutrit is transmitted through an effectively two-dimensional quantum channel.

To quantify the visual agreement between reconstructed and simulated matrices, we use the matrix distance
\begin{equation}
D(\rho_{\rm exp},\rho_{\rm sim})
=\sqrt{{\rm Tr}\left[\left(\rho_{\rm exp}-\rho_{\rm sim}\right)^2\right]} .
\label{eq:supp_matrix_distance}
\end{equation}
This follows the matrix-distance convention used for microwave-link process comparisons in Ref.~\cite{supp:kurpiers2018deterministic}. It is zero for identical matrices and increases as the reconstructed matrix moves away from the simulation. For the qutrit-transfer process matrix in main-text Fig.~3, the mean reconstructed matrix and the simulated matrix give \(D=0.141\).

For REG, the target state is
\begin{equation}
\ket{\psi_{\rm REG}}=\frac{1}{\sqrt{3}}\left(\ket{gf}+\ket{ee}+\ket{fg}\right),
\label{eq:supp_reg_target}
\end{equation}
up to local phases. The raw density matrix includes deterministic phases accumulated during the pulse sequence. Before evaluating the phase-aligned fidelity and dense-coding capacity, we remove these phases by local virtual-Z rotations. The extracted \(ef\)-frame corrections are \(0.911~{\rm rad}\) on node A and \(-2.035~{\rm rad}\) on node B.

After phase alignment, the REG fidelity is \(81.21\%\pm0.37\%\). The reconstructed matrix retains the three target populations and the three pairwise coherences in the \(\{\ket{gf},\ket{ee},\ket{fg}\}\) manifold. The corresponding simulation gives a REG fidelity of \(83.36\%\), using the same loss and dephasing parameters as the qutrit-transfer simulation.

The imaginary components in Fig.~\ref{fig:supp_tomography_phase_diagnostics}(a,b) remain small compared with the dominant real coherences. The residual maps in Fig.~\ref{fig:supp_tomography_phase_diagnostics}(c,d) further show that the remaining experiment--simulation differences are distributed over small off-diagonal and leakage-related components rather than a single large coherent phase error. This supports the interpretation that the reconstructed qutrit transfer and REG state are mainly limited by finite transfer efficiency, internal loss and dephasing.

\begin{table}[htbp]
\centering
\caption{\textbf{Matrix-distance comparison between experiment and pulse-level simulation.}
The mean-matrix distance is calculated between the readout-corrected mean experimental matrix and the simulated mean matrix. The repetition-level value is the mean and standard deviation of the distance between each experimental tomography repetition and the same simulated mean matrix.}
\label{tab:supp_exp_sim_matrix_distance}
\begin{tabular}{llll}
\hline
Figure & Matrix & \(D\) & Repetition-level \(D\) \\
\hline
Main-text Fig.~2 & \(e0g1\) primitive process matrix & 0.029 & \(0.039\pm0.014\) \\
Main-text Fig.~2 & \(f0g1\) primitive process matrix & 0.068 & \(0.085\pm0.016\) \\
Main-text Fig.~3 & qutrit-transfer process matrix & 0.141 & \(0.142\pm0.002\) \\
Main-text Fig.~4 & REG density matrix & 0.119 & \(0.120\pm0.003\) \\
\hline
\end{tabular}
\end{table}

\section{\MakeUppercase{Beyond-qubit benchmarks and robustness checks}}
\label{sec:supp_high_dimensional_benchmarks_confidence}
\label{sec:supp_beyond_qubit_benchmarks_confidence}

\subsection{Metric definitions and readout-correction confidence}
\label{sec:supp_high_dimensional_metrics}

\begin{figure}[htbp]
    \centering
    \suppfiginclude{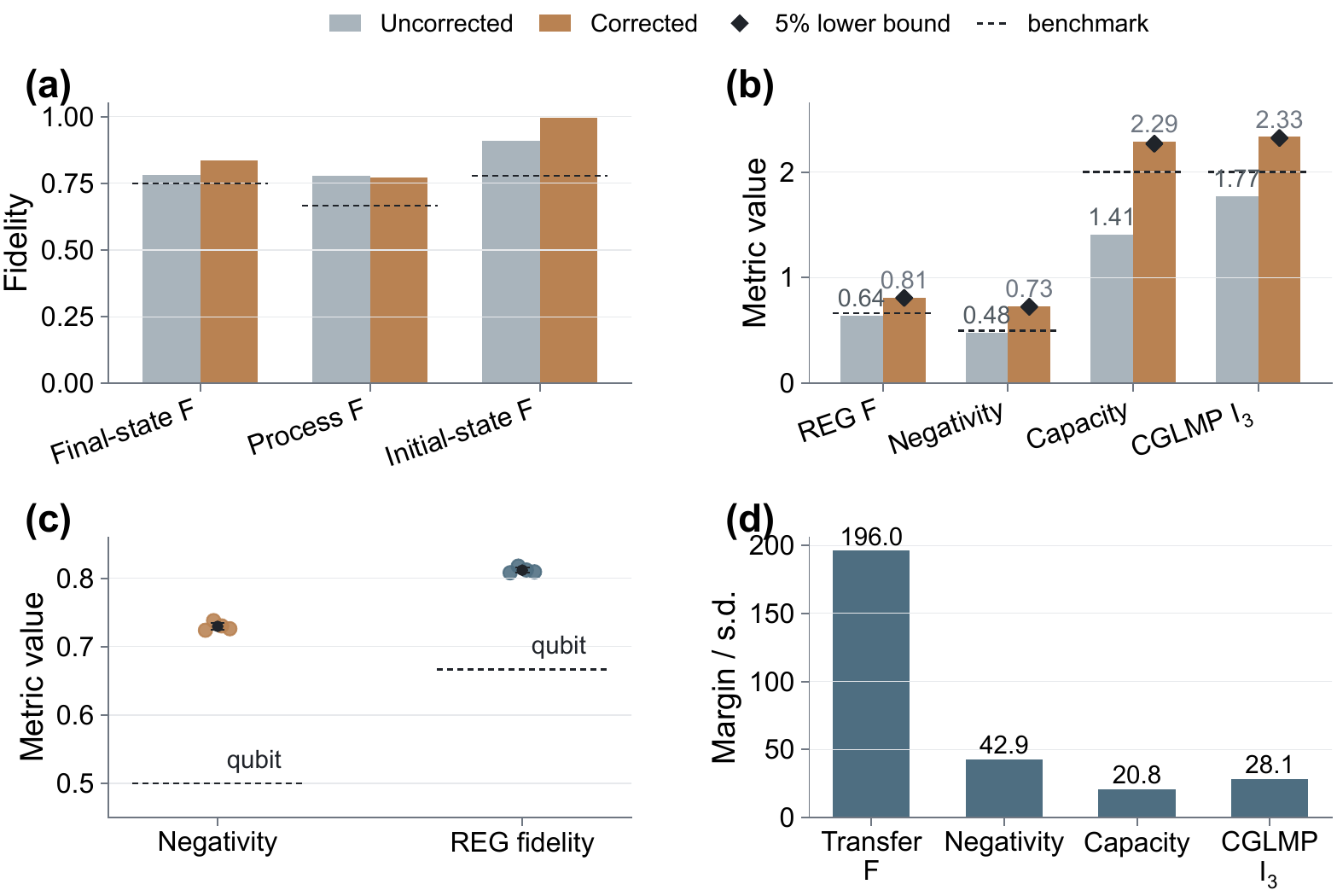}{width=0.95\textwidth}
    \caption{\textbf{Readout-correction sensitivity and repetition-level confidence.}
    \textbf{(a)} Transfer-state and process fidelities reconstructed with and without the independently calibrated readout-assignment correction. Short dashed segments mark the relevant bounds for the three different quantities.
    \textbf{(b)} REG metrics reconstructed with and without the same correction. Black diamonds mark the 5th-percentile lower bounds obtained by Monte-Carlo resampling of the calibrated readout-assignment matrices.
    \textbf{(c)} Repetition-level spread of the readout-corrected REG negativity and phase-aligned REG fidelity. Dots are individual tomography repetitions, black markers show the mean and one standard deviation, and dashed segments mark the two-qubit bounds.
    \textbf{(d)} Statistical margin above the relevant qubit or local benchmark, normalized by the repetition-level standard deviation.}
    \label{fig:supp_reg_information_metrics}
\end{figure}

Figure~\ref{fig:supp_reg_information_metrics} is the main robustness check for the beyond-qubit claims. The quantities plotted there are the Schmidt-number witness~\cite{supp:terhal2000schmidt}, negativity~\cite{supp:peres1996separability,supp:vidal2002computable,supp:horodecki2009quantum}, tomography-inferred dense-coding capacity~\cite{supp:bennett1992communication,supp:horodecki2001classical,supp:bowen2001classical}, and tomography-inferred CGLMP parameter~\cite{supp:collins2002bell}. The logic is deliberately conservative. We first compare corrected and detector-convolved reconstructions to show the size of the readout bias. We then resample the independently calibrated assignment matrices and repeat the full reconstruction. The main-text claims use the readout-corrected tomography with the measured assignment matrices in Fig.~\ref{fig:supp_readout_correction}; the black diamonds in Fig.~\ref{fig:supp_reg_information_metrics} show that the relevant conclusions remain above their limits under this assignment-matrix uncertainty.

In Fig.~\ref{fig:supp_reg_information_metrics}(a), we compare the reconstructed final-state fidelity, the qutrit process fidelity, and the input-state fidelity before and after applying the calibrated readout-assignment correction. The mean final-state fidelity is \(83.68\%\) with readout correction and \(78.26\%\) without it. The final-state fidelity is the mean fidelity of the transferred qutrit states at the receiver, and its dashed segment marks the average-fidelity benchmark \(3/4\) for transferring an arbitrary qutrit through an effectively two-dimensional quantum channel. The qutrit process fidelity has a different normalization. Using \(F_{\rm avg}=(dF_{\rm pro}+1)/(d+1)\) with \(d=3\), the same \(F_{\rm avg}=3/4\) benchmark corresponds to \(F_{\rm pro}=2/3\), shown as the dashed segment for the process-fidelity group. The input-state fidelity characterizes preparation of the nine QPT input states. For this finite input set, the best average fidelity obtainable with a fixed two-dimensional qutrit subspace is \(7/9\), obtained from the two largest eigenvalues of the average ideal input-state density matrix. This finite-set bound is shown for the input-state fidelity group.

In Fig.~\ref{fig:supp_reg_information_metrics}(b), we make a comparison between four REG-derived quantities before and after readout correction in their native units: the phase-aligned target-state overlap, the negativity, the tomography-inferred dense-coding capacity and the tomography-inferred CGLMP value \(I_3\). The dashed segments mark the corresponding bounds: \(2/3\) for the REG fidelity, \(0.5\) for the two-qubit negativity bound, \(2\) bits for the two-qubit dense-coding limit, and \(I_3=2\) for locality. To test sensitivity to the calibrated assignment matrices, we resample the two qutrit readout matrices with a conservative Dirichlet model using \(4000\) effective calibration shots per prepared level and a \(0.3\%\) symmetric drift floor. This drift floor accounts for slow readout-response variations not captured by finite-shot Dirichlet sampling; it is applied symmetrically over the three possible outcomes and is not fitted to the reconstructed density matrix or to any benchmark value. Its magnitude is chosen as a conservative sub-percent broadening of the independently calibrated assignment matrices. The black diamonds are the 5th-percentile lower bounds from \(400\) resampled reconstructions. Even the worst sampled reconstruction gives \(F_{\Psi_3}=0.8036\), \({\cal N}=0.7176\), \(C=2.242\) bits and \(I_3=2.307\), all above the corresponding bounds. The beyond-qubit conclusions thus persist across the sampled readout matrices.

Fig.~\ref{fig:supp_reg_information_metrics}(c) shows the repetition-level spread of the two central REG quantities after readout correction: negativity and phase-aligned REG fidelity. This repetition-level comparison checks the consistency of the excess above the two-qubit bounds.

Fig.~\ref{fig:supp_reg_information_metrics}(d) plots the margin \((\bar{x}-x_{\rm bound})/\sigma_x\) for four benchmark quantities. This is a compact confidence diagnostic based on repetition-level spread. Systematic effects from readout calibration, pulse calibration and tomography reconstruction are treated separately through the readout-correction comparison and the simulation model.

For the REG-state fidelity, the relevant two-level entanglement bound is \(2/3\) because for a target maximally entangled qutrit state with three equal Schmidt coefficients, the maximum overlap with any Schmidt-rank-two state is \(2/3\). The measured phase-aligned fidelity \(F_{\Psi_3}=0.8121\pm0.0037\) therefore witnesses Schmidt number \(3\). Under the readout-assignment resampling described above, the 5th-percentile lower bound is \(F_{\Psi_3}=0.8086\), and the worst sampled value is \(F_{\Psi_3}=0.8036\), still well above \(2/3\). This is the qubit-entanglement bound used in Fig.~\ref{fig:supp_reg_information_metrics}(c).

The negativity is computed from the partial transpose~\cite{supp:peres1996separability,supp:vidal2002computable},
\begin{equation}
{\cal N}(\rho)=\frac{\|\rho^{T_B}\|_1-1}{2}.
\label{eq:supp_negativity}
\end{equation}
For a maximally entangled two-qubit state, the maximum negativity is \(0.5\). The measured ensemble-averaged REG state has \({\cal N}=0.7297\pm0.0054\), exceeding the two-qubit bound and confirming that the generated entanglement cannot be represented as a maximally entangled two-level pair. The log-negativity is \(1.2984\pm0.0063\). The readout-assignment resampling gives a 5th-percentile lower bound \({\cal N}=0.7248\) and a worst sampled value \({\cal N}=0.7176\), again above the two-qubit maximum.

Following the standard dense-coding capacity expression for a shared bipartite state~\cite{supp:bennett1992communication,supp:bowen2001classical}, the dense-coding capacity is evaluated from the reconstructed shared state as
\begin{equation}
C=\log_2 d + S(\rho_B)-S(\rho_{AB}),
\label{eq:supp_dense_coding_capacity}
\end{equation}
with local dimension \(d=3\). The measured phase-aligned ensemble-averaged density matrix gives \(C=2.273\pm0.014\) bits. This value exceeds both the classical qutrit communication limit \(\log_2 3\) and the two-qubit dense-coding limit of \(2\) bits, demonstrating the channel-capacity advantage provided by the high-dimensional entangled resource. The corresponding 5th-percentile lower bound under readout-assignment resampling is \(C=2.267\) bits, and the worst sampled value is \(C=2.242\) bits.

Finally, we evaluate the Collins-Gisin-Linden-Massar-Popescu (CGLMP) parameter \(I_3\) from the reconstructed two-qutrit density matrix~\cite{supp:collins2002bell}. The CGLMP inequality is the standard Bell inequality for two parties, two measurement settings per party, and three outcomes per measurement. We first express the physical target state \((\ket{gf}+\ket{ee}+\ket{fg})/\sqrt{3}\) in the logical basis \((\ket{00}+\ket{11}+\ket{22})/\sqrt{3}\) by reversing the level order of node B. In this logical basis, Alice chooses one of two measurements \(A_0,A_1\), Bob chooses one of two measurements \(B_0,B_1\), and all outcomes are evaluated modulo 3. We use
\begin{align}
I_3=&\,P(A_0-B_0=0)+P(B_0-A_1=1)+P(A_1-B_1=0)+P(B_1-A_0=0) \nonumber\\
&-P(A_0-B_0=-1)-P(B_0-A_1=0)-P(A_1-B_1=-1)-P(B_1-A_0=-1),
\label{eq:supp_cglmp_i3}
\end{align}
where \(P(A_x-B_y=k)=\sum_{a=0}^{2}P(a,b=a-k\,{\rm mod}\,3|x,y)\). Any local hidden-variable model satisfies \(I_3\leq2\).

The projective measurement bases are the usual CGLMP bases optimized for the maximally entangled qutrit state,
\begin{align}
\ket{a_x} &= \frac{1}{\sqrt{3}}\sum_{j=0}^{2}
\omega^{j(a+\alpha_x)}\ket{j},&
\ket{b_y} &= \frac{1}{\sqrt{3}}\sum_{j=0}^{2}
\omega^{j(-b+\beta_y)}\ket{j},
\label{eq:supp_cglmp_bases}
\end{align}
with \(\omega=\exp(2\pi i/3)\), \((\alpha_0,\alpha_1)=(0,1/2)\), and \((\beta_0,\beta_1)=(1/4,-1/4)\). For a density matrix \(\rho\), the probabilities are computed as
\begin{equation}
P(a,b|x,y)=
\bra{a_x,b_y}\rho\ket{a_x,b_y}.
\label{eq:supp_cglmp_prob}
\end{equation}
With this convention, the ideal maximally entangled qutrit target gives \(I_3=2.873\), while the local realistic bound remains \(2\). The reconstructed state gives \(I_3=2.332\pm0.012\), above the local bound and consistent with nonlocal correlations available from the generated two-qutrit entangled state. The 5th-percentile lower bound under readout-assignment resampling is \(I_3=2.322\), and the worst sampled value is \(I_3=2.307\). Since this quantity is inferred from the reconstructed density matrix rather than obtained from a direct loophole-free Bell experiment with independently chosen measurement settings, we describe it as a tomography-inferred CGLMP violation.

Together, these metrics show the same high-dimensional advantage from three complementary viewpoints: the entanglement negativity exceeds the two-qubit maximum, the tomography-inferred dense-coding capacity exceeds the qubit limit, and the tomography-inferred correlations exceed the qutrit CGLMP local bound. The readout-resampling test in Fig.~\ref{fig:supp_reg_information_metrics} is the key systematic check supporting these statements, because it propagates uncertainty in the independently measured qutrit assignment matrices through the same reconstruction pipeline used for the main-text numbers.

\section{\MakeUppercase{Pulse-level simulation, dephasing model and error budget}}
\label{sec:supp_simulation_waveform_choice}
\label{sec:supp_pulse_level_simulation_waveform_choice}

\subsection{Cascaded master-equation model and dephasing implementation}
\label{sec:supp_cascaded_model_dephasing}

The simulations use the same pulse-level cascaded model used to reproduce the primitive-calibration data. Each node is represented by a truncated transmission resonator \(a_j\) and a truncated transmon mode \(b_j\), with \(j=A,B\). The tensor order is \((a_A,q_A,a_B,q_B)\), and the transmon Hilbert space is restricted to the qutrit manifold \(\{\ket{g},\ket{e},\ket{f}\}\). Protocol-level parameters are set by the intended operation: full-transfer pulses implement the two qutrit-transfer primitives, whereas reduced pulse areas implement the partial transfers used for remote entanglement generation. The chirp calibrations described in Sec.~\ref{sec:supp_chirped_pulse_calibration} are treated as part of the calibrated drive frame. We therefore set the residual deterministic detunings of the \(e0g1\) and \(f0g1\) pulses to zero and do not add separate ac-Stark-shift correction terms in the rotating-frame Hamiltonian. For one quasi-static detuning realization, the Hamiltonian is
\begin{align}
H(t)/\hbar
=H_{\rm stat}/\hbar+H_{\rm casc}/\hbar
+H_{e0g1}(t)/\hbar+H_{f0g1}(t)/\hbar
+H_{\rm loc}(t)/\hbar+H_{\rm noise}(\delta)/\hbar .
\label{eq:supp_total_hamiltonian}
\end{align}
The static part keeps the weak auxiliary-resonator Kerr terms and the dispersive shifts that are present in the pulse-level notebooks,
\begin{align}
H_{\rm stat}/\hbar
=\sum_{j=A,B}\left[
\frac{K_j}{2}a_j^{\dagger 2}a_j^2
+\chi_j a_j^\dagger a_j b_j^\dagger b_j
\right],
\label{eq:supp_static_hamiltonian}
\end{align}
with the residual resonator detuning set to zero at the matched operating point. The directional waveguide connection is written in the Gardiner--Carmichael cascaded form
\begin{equation}
H_{\rm casc}/\hbar
=-\frac{i}{2}\sqrt{\eta_c\kappa_{T,A}\kappa_{T,B}}
\left(a_Aa_B^\dagger-a_A^\dagger a_B\right),
\label{eq:supp_cascaded_hamiltonian}
\end{equation}
where \(\kappa_{T,A}/2\pi=\kappa_{T,B}/2\pi\simeq10~{\rm MHz}\) is the loaded auxiliary-resonator linewidth at the high-bandwidth Purcell-filter setting and \(\eta_c=0.92\) is the channel efficiency used in the main-text simulations.

The two transfer primitives use distinct effective interactions. For the \(e0g1\) primitive, flux modulation activates the resonant exchange
\begin{align}
H_{e0g1}(t)/\hbar
=\sum_{j=A,B} g_{e,j}(t)
\left(\ket{0,e}_j\bra{1,g}+\ket{1,g}_j\bra{0,e}\right),
\label{eq:supp_e0g1_hamiltonian}
\end{align}
which transfers \(\ket{e,0}\leftrightarrow\ket{g,1}\) between the qutrit and the auxiliary resonator. For the \(f0g1\) primitive, the cavity-assisted Raman process is modeled by the three-wave-mixing effective term used in the original simulation,
\begin{align}
H_{f0g1}(t)/\hbar
=\sum_{j=A,B}\frac{g_{f,j}(t)}{\sqrt{2}}
\left(b_j^{\dagger 2}a_j+a_j^\dagger b_j^2\right),
\label{eq:supp_f0g1_hamiltonian}
\end{align}
which couples \(\ket{f,0}\leftrightarrow\ket{g,1}\) in the qutrit subspace. In the experiments analyzed here, \(g_{e,j}(t)\) and \(g_{f,j}(t)\) are chirp-corrected \(200~{\rm ns}\) sine envelopes. Table~\ref{tab:supp_pulse_level_coupling_parameters} lists the effective peak couplings used in the pulse-level simulations.

\begin{table}[htbp]
\centering
\caption{\textbf{Effective coupling parameters used in the pulse-level simulations for Figs.~2--4.}
All rows use chirp-corrected \(200~{\rm ns}\) sine envelopes for the addressed photon-mediated primitive. The listed values are the peak resonant-frame couplings \(g_j^{\rm max}/2\pi\) used in the cascaded master-equation simulations. The Fig.~3 qutrit-transfer sequence applies \(f0g1\) followed by \(e0g1\), whereas the Fig.~4 REG sequence applies \(e0g1\) followed by \(f0g1\).}
\label{tab:supp_pulse_level_coupling_parameters}
\small
\begin{tabular}{lllcc}
\hline
Use & Primitive & Target operation & \(g_A^{\rm max}/2\pi\) & \(g_B^{\rm max}/2\pi\) \\
\hline
Fig.~2 dynamics & \(e0g1\) & full transfer & \(3.90~{\rm MHz}\) & \(3.90~{\rm MHz}\) \\
Fig.~2 dynamics & \(f0g1\) & full transfer & \(3.85~{\rm MHz}\) & \(3.35~{\rm MHz}\) \\
Fig.~3 QPT & \(f0g1\) & \(\pi\) transfer & \(4.00~{\rm MHz}\) & \(3.45~{\rm MHz}\) \\
Fig.~3 QPT & \(e0g1\) & \(\pi\) transfer & \(3.90~{\rm MHz}\) & \(3.90~{\rm MHz}\) \\
Fig.~4 REG & \(e0g1\) & partial, \(P=1/3\) & \(1.31~{\rm MHz}\) & \(3.92~{\rm MHz}\) \\
Fig.~4 REG & \(f0g1\) & partial, \(P=1/2\) & \(1.78~{\rm MHz}\) & \(4.25~{\rm MHz}\) \\
\hline
\end{tabular}
\end{table}

The local-control Hamiltonian contains only the qutrit rotations used in the protocols,
\begin{align}
H_{\rm loc}(t)/\hbar
=\sum_{j=A,B}\left[
\frac{\Omega_{ge,j}(t)}{2}\left(\ket{g_j}\bra{e_j}+{\rm h.c.}\right)
+\frac{\Omega_{ef,j}(t)}{2}\left(\ket{e_j}\bra{f_j}+{\rm h.c.}\right)
\right],
\label{eq:supp_local_hamiltonian}
\end{align}
with Gaussian \(\pi_{ge}\), \(\pi/2_{ge}\), \(\pi_{ef}\), and \(\pi/2_{ef}\) pulses inserted according to the state-transfer or REG pulse sequence.

The simulations are integrated with QuTiP~\cite{supp:JOHANSSON20121760,supp:JOHANSSON20131234}. In the implementation, Eq.~\eqref{eq:supp_cascaded_hamiltonian} is combined with the collective radiative collapse operator
\begin{equation}
C_{\rm casc}
=\sqrt{\eta_c\kappa_{T,A}}\,a_A+\sqrt{\kappa_{T,B}}\,a_B,
\label{eq:supp_collective_collapse}
\end{equation}
and the channel-loss operator
\begin{equation}
C_{\rm loss,A}=\sqrt{(1-\eta_c)\kappa_{T,A}}\,a_A .
\label{eq:supp_channel_loss}
\end{equation}
This representation is equivalent to the usual unidirectional cascaded master equation and enforces propagation from node A to node B. For a fixed detuning realization,
\begin{align}
\frac{d\rho}{dt}
=-i[H(t),\rho]
+{\cal D}[C_{\rm casc}]\rho
+{\cal D}[C_{\rm loss,A}]\rho
+\sum_m{\cal D}[C_m]\rho ,
\label{eq:supp_master_equation}
\end{align}
where
\begin{equation}
{\cal D}[C]\rho=C\rho C^\dagger-\frac{1}{2}\{C^\dagger C,\rho\}.
\label{eq:supp_lindblad}
\end{equation}
The remaining \(C_m\) describe internal auxiliary-resonator loss and qutrit relaxation,
\begin{align}
C_{{\rm int},j}&=\sqrt{\kappa_{{\rm int},j}}a_j, \notag\\
C_{ge,j}&=\sqrt{1/T_{1,ge,j}}\ket{g_j}\bra{e_j},\qquad
C_{ef,j}=\sqrt{1/T_{1,ef,j}}\ket{e_j}\bra{f_j}.
\label{eq:supp_loss_collapse}
\end{align}
The intrinsic auxiliary-resonator losses used in the present simulations are \(\kappa_{{\rm int},A}/2\pi=0.0208~{\rm MHz}\) and \(\kappa_{{\rm int},B}/2\pi=0.0176~{\rm MHz}\), corresponding to the angular-frequency rates \(0.1308~\mu{\rm s}^{-1}\) and \(0.1104~\mu{\rm s}^{-1}\). The independently measured \(T_{1,ge}\) and \(T_{1,ef}\) values are listed in Fig.~\ref{fig:supp_device_coherence_readout_simulation}.

The main simulations represent pure dephasing through quasi-static \(1/f\)-like detuning noise. This choice is motivated by the Ramsey data: after the relaxation-limited decay is fixed by \(T_1\), the remaining envelope is closer to a Gaussian decay than to a single exponential decay. A Markovian pure-dephasing model is useful as a comparison, but it continuously removes phase information during every part of the sequence and therefore over-penalizes the multi-pulse qutrit-transfer and REG protocols.

For one realization, we sample
\begin{equation}
\delta=(\delta_{ge,A},\delta_{ef,A},\delta_{ge,B},\delta_{ef,B}),
\qquad
\delta_\mu\sim{\cal N}(0,\sigma_\mu^2),
\label{eq:supp_static_detuning_distribution}
\end{equation}
and add the static detuning Hamiltonian
\begin{align}
H_{\rm noise}(\delta)/\hbar
=\sum_{j=A,B}\left[
\delta_{ge,j}\left(\ket{e_j}\bra{e_j}+\ket{f_j}\bra{f_j}\right)
+\delta_{ef,j}\ket{f_j}\bra{f_j}
\right].
\label{eq:supp_noise_hamiltonian}
\end{align}
With this convention, the \(ge\) coherence accumulates \(\delta_{ge,j}\), while the \(ef\) coherence accumulates \(\delta_{ef,j}\). The final population traces, process matrices and density matrices are obtained by averaging the simulated output density matrix over the quasi-static ensemble, using the same tomography objects as in the experimental analysis.

The detuning widths are obtained from the \(T_1\)-corrected Ramsey fits in Eq.~\eqref{eq:supp_ramsey_fit}: \(\sigma_{ge,A}/2\pi=59.24~{\rm kHz}\), \(\sigma_{ge,B}/2\pi=102.47~{\rm kHz}\), \(\sigma_{ef,A}/2\pi=60.01~{\rm kHz}\), and \(\sigma_{ef,B}/2\pi=102.18~{\rm kHz}\). These experimentally extracted quasi-static detuning widths are used directly in the simulations of the primitive transfer dynamics, qutrit process tomography and two-qutrit state tomography, so that the same dephasing model is applied consistently to Figs.~2--4.

The model is nevertheless a fixed-parameter description of a long acquisition. The population-dynamics traces in Fig.~2 are acquired on a shorter time scale than the full qutrit QPT and two-qutrit QST datasets, whereas the tomography experiments average over many preparation, analysis and readout settings. During this longer acquisition, the two node frequencies, the \(e0g1\) and \(f0g1\) resonance conditions, the Purcell-filter operating point, the readout assignment matrices and the local phase reference can drift slowly. These drifts are not explicitly sampled in the pulse-level simulation, which uses one set of calibrated pulse amplitudes, resonant-frame couplings, readout-independent output states and virtual-\(Z\) phases. They can therefore contribute to the residual experiment--simulation differences in Figs.~3 and 4, especially in off-diagonal coherences and tomography-derived benchmarks, without changing the short-time transfer dynamics as strongly.

\begin{figure}[htbp]
    \centering
    \suppfiginclude{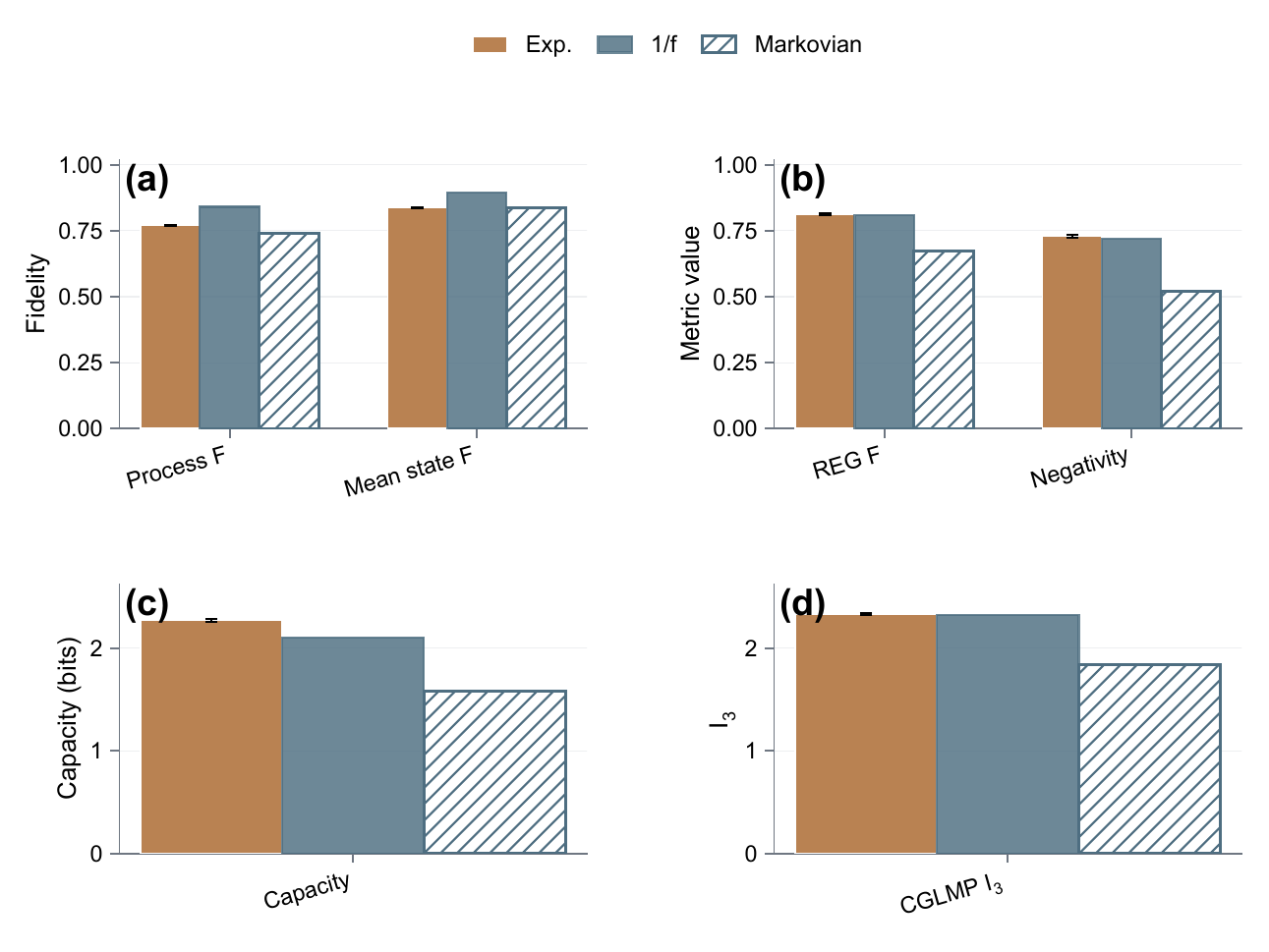}{width=0.95\textwidth}
    \caption{\textbf{Comparison between Markovian and quasi-static dephasing models.}
    \textbf{(a)} Qutrit state-transfer process fidelity and mean output-state fidelity.
    \textbf{(b)} REG state fidelity and negativity, plotted on a \(0\)--\(1\) scale.
    \textbf{(c)} Dense-coding capacity inferred from the REG density matrix.
    \textbf{(d)} Tomography-inferred CGLMP value \(I_3\) from the same density matrix.}
    \label{fig:supp_dephasing_model_comparison}
\end{figure}

Figure~\ref{fig:supp_dephasing_model_comparison} compares this quasi-static model with a Markovian pure-dephasing model using the same pulse-level settings as the main-text qutrit-transfer and REG simulations. The channel efficiency, internal resonator loss, pulse amplitudes and pulse timing are kept fixed. Only the dephasing implementation is changed. For the Markovian comparison, the Ramsey traces are refitted with an exponential envelope. After subtracting the relaxation-limited coherence decay, the resulting pure-dephasing times are \(T_{\phi,ge,A}^{\rm M}=3.84~\mu{\rm s}\), \(T_{\phi,ge,B}^{\rm M}=1.71~\mu{\rm s}\), \(T_{\phi,ef,A}^{\rm M}=3.79~\mu{\rm s}\), and \(T_{\phi,ef,B}^{\rm M}=1.85~\mu{\rm s}\). These times set the Lindblad dephasing rates used only for this comparison. In the quasi-static model, the dominant low-frequency detuning is constant during one experimental shot and changes between shots. This produces the measured Gaussian Ramsey envelope, but it does not erase coherence inside a \(200~{\rm ns}\) transfer pulse as strongly as the Markovian model. The difference between the two descriptions is modest for the qutrit state-transfer benchmarks: the Markovian simulation gives a process fidelity of \(0.739\) and a mean output-state fidelity of \(0.839\), compared with \(0.841\) and \(0.895\) from the quasi-static \(1/f\)-like simulation. The difference is larger for REG, where the Markovian simulation gives a state fidelity of \(0.728\), negativity of \(0.600\), tomography-inferred dense-coding capacity of \(1.791\) bits and a tomography-inferred CGLMP value \(I_3=2.052\), whereas the quasi-static simulation gives \(0.834\), \(0.756\), \(2.242\) bits and \(I_3=2.417\) after first averaging the simulated density matrices. Thus, for qutrit state transfer the Markovian conversion changes the benchmarks only moderately, whereas for REG it suppresses the coherences that set the negativity, tomography-inferred dense-coding capacity and tomography-inferred CGLMP value much more strongly.

\begin{table}[htbp]
    \centering
    \caption{\textbf{Markovian dephasing inputs and numerical comparison of dephasing models.}
    The first table compares the pure-dephasing inputs extracted from the same Ramsey and \(T_1\) data under two assumptions. \(T_\phi^{\rm M}\) is obtained from an exponential Ramsey envelope after subtracting the relaxation-limited coherence decay. \(T_\phi^{1/f}\equiv\sqrt{2}/\sigma\) is obtained from the \(T_1\)-corrected Gaussian Ramsey envelope used for the quasi-static \(1/f\)-like detuning ensemble. Experimental benchmark values include one-standard-deviation statistical uncertainties. The two simulations use the same pulse-level parameters and differ only in the dephasing model.}
    \label{tab:supp_dephasing_model_comparison}
    \begin{tabular}{lcccc}
        \hline
        Channel & \(T_2^{\rm M}\) (\(\mu{\rm s}\)) & \(T_\phi^{\rm M}\) (\(\mu{\rm s}\)) & \(\sigma/2\pi\) (kHz) & \(T_\phi^{1/f}\) (\(\mu{\rm s}\)) \\
        \hline
        \(ge,A\) & \(3.528\) & \(3.839\) & \(59.24\) & \(3.80\) \\
        \(ge,B\) & \(1.647\) & \(1.711\) & \(102.47\) & \(2.20\) \\
        \(ef,A\) & \(2.958\) & \(3.787\) & \(60.01\) & \(3.75\) \\
        \(ef,B\) & \(1.631\) & \(1.854\) & \(102.18\) & \(2.20\) \\
        \hline
    \end{tabular}

    \vspace{0.8em}

    \begin{tabular}{lccc}
        \hline
        Benchmark & Experiment & Quasi-static \(1/f\) & Markovian \\
        \hline
        QST process fidelity & \(0.771\pm0.001\) & \(0.841\) & \(0.739\) \\
        QST mean state fidelity & \(0.837\pm0.0004\) & \(0.895\) & \(0.839\) \\
        REG fidelity & \(0.812\pm0.004\) & \(0.834\) & \(0.728\) \\
        Negativity & \(0.730\pm0.005\) & \(0.756\) & \(0.600\) \\
        Dense-coding capacity (bits) & \(2.273\pm0.014\) & \(2.242\) & \(1.791\) \\
        CGLMP \(I_3\) & \(2.332\pm0.012\) & \(2.417\) & \(2.052\) \\
        \hline
    \end{tabular}
\end{table}

\subsection{Sine and sech waveform comparison}
\label{sec:supp_sine_sech_waveform}

\begin{figure}[htbp]
    \centering
    \suppfiginclude{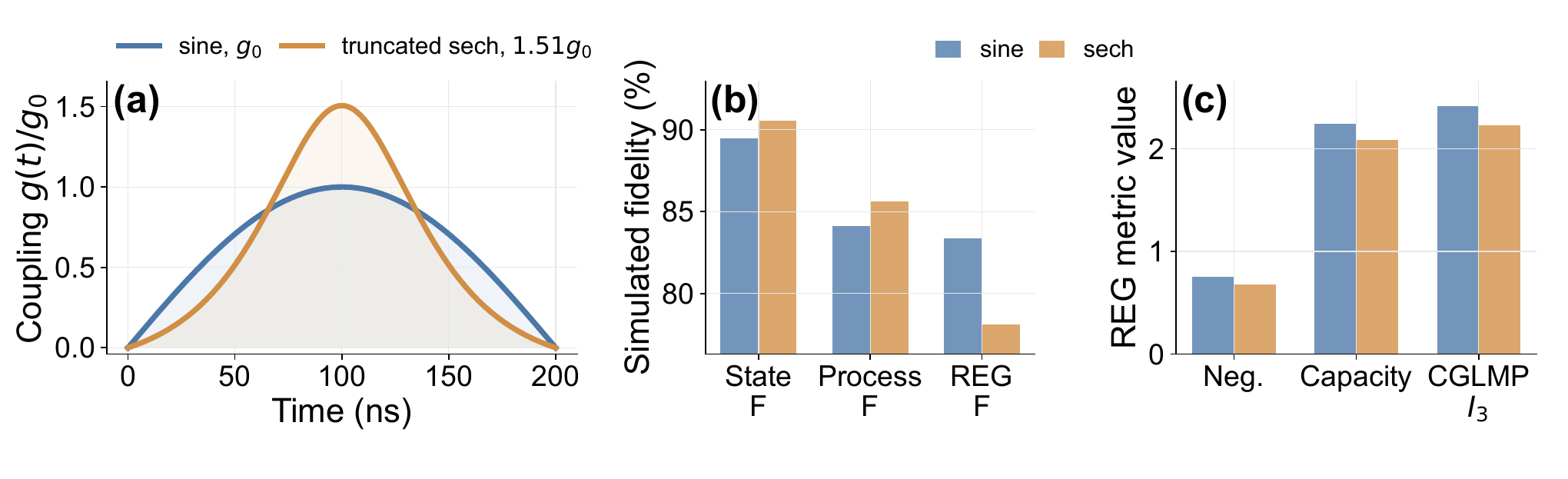}{width=0.95\textwidth}
    \caption{\textbf{Waveform choice and linewidth-set sech-pulse comparison.}
    \textbf{(a)} The \(200~{\rm ns}\) sine control pulse used in the experiment and a finite-window sech pulse derived from \(g(t)=g_0/\cosh[(t-t_0)/\tau]\). The sech scale is set by \(\kappa_{\rm sech}/2\pi=5~{\rm MHz}\), motivated by the matched transmission-resonator linewidth \(\kappa/2\pi\simeq10~{\rm MHz}\). The endpoint value of the sech window is subtracted before area matching so that the pulse returns to zero at the experimental pulse boundaries. Matching the pulse area requires a larger sech peak coupling, \(g_{\rm sech}^{\rm max}\simeq1.51g_0\).
    \textbf{(b)} Qutrit state-transfer fidelity, qutrit process fidelity and REG state fidelity using the experimental sine envelope and the linewidth-set finite-window sech envelope.
    \textbf{(c)} REG information metrics from the same decoherence model: negativity, tomography-inferred dense-coding capacity and tomography-inferred CGLMP value.}
    \label{fig:supp_sine_sech_waveform}
\end{figure}

\begin{table}[htbp]
\centering
\caption{\textbf{Pulse-level comparison of sine and finite-window sech waveforms.}
The finite-window sech is constrained to the same \(200~{\rm ns}\) pulse window as the experimental sine pulse and is area matched to the sine pulse. Both rows use the same loss, \(T_1\), \(T_\phi\), \(\eta_c\), \(\kappa_{\rm int}\), quasi-static detuning ensemble and REG averaged-density-matrix convention as the main-text simulations.}
\label{tab:supp_waveform_decoherence_comparison}
\begin{tabular}{llcccccc}
\hline
Waveform & Model & QST state F & QST process F & REG F & \(N\) & \(C\) (bits) & \(I_3\) \\
\hline
sine & with decoherence & 0.8945 & 0.8410 & 0.8336 & 0.7556 & 2.2424 & 2.4172 \\
finite-window sech & with decoherence & 0.9056 & 0.8561 & 0.7809 & 0.6822 & 2.0871 & 2.2328 \\
\hline
\end{tabular}
\end{table}

The transfer experiments in the main text use \(200~{\rm ns}\) sine-shaped parametric pulses because they produce a near time-symmetric photon temporal mode for the calibrated bandwidth, are experimentally robust, have low spectral leakage, and can be applied reproducibly to both the \(e0g1\) and \(f0g1\) primitives. Pulse-shape optimization can further improve photon capture. In the ideal tunable-coupler problem, near-unit state-transfer efficiency is obtained by shaping the emission and capture rates so that the receiver implements the time reverse of the emitted photon mode~\cite{supp:korotkov2011flying,supp:wenner2014catching}. Related superconducting-circuit experiments have also used microwave control to generate and symmetrize shaped photons~\cite{supp:pechal2014microwave,supp:srinivasan2014time}. Following the pulse definition used in our original simulations, the sech envelope is written as \(g(t)=g_0/\cosh[(t-t_0)/\tau]\).

The sine-pulse simulations reproduce the calibrated experimental protocol. We include the sech comparison in Fig.~\ref{fig:supp_sine_sech_waveform} as a diagnostic of how much performance is left in pulse-shape optimization for a full emission-and-capture process. The first comparison keeps the same \(200~{\rm ns}\) pulse window as the experiment and matches the integrated coupling area. Since the matched transmission-resonator linewidth is \(\kappa/2\pi\simeq10~{\rm MHz}\), we use a reference sech scale \(\kappa_{\rm sech}/2\pi=5~{\rm MHz}\), corresponding to \(\tau=1/\kappa_{\rm sech}\simeq31.8~{\rm ns}\). Starting from the original simulation definition \(g(t)=g_0/\cosh[(t-t_0)/\tau]\), we subtract the finite-window endpoint value and renormalize the peak before area matching, so that the plotted sech pulse starts and ends continuously at zero. Under this constraint the peak coupling is \(g_{\rm sech}^{\rm max}\simeq1.51g_0\), larger than the sine-pulse peak.

The presence of the tunable Purcell-filter interface makes this optimization slightly different from the textbook single-resonator time-reversal problem. The emitted field is filtered by an additional mode before entering the transmission line, so the photon mode seen by the receiver is not exactly the mode generated by the auxiliary resonator alone. A receiver pulse designed for the unfiltered single-pole response is therefore not exactly optimal once the Purcell filter is included. For parameters close to the present device, this mode mismatch gives an order-\(1\%\) capture-efficiency penalty. The numerical values are summarized in Table~\ref{tab:supp_waveform_decoherence_comparison}. With the current decoherence model, the finite-window sech pulse slightly improves the full qutrit-transfer state fidelity relative to the sine pulse.

The REG trend is different. Directly replacing the sine pulses by the finite-window sech envelope lowers the reconstructed REG benchmarks in the present calculation, even though the same waveform improves the full-transfer channel. This comparison tests a fixed pulse substitution in the calibrated sequence. REG relies on partial-transfer interference, so its performance depends on the relative amplitudes and phases of the \(\ket{gf}\), \(\ket{ee}\), and \(\ket{fg}\) components and on chirp compensation during the partial pulses. Optimizing REG therefore requires joint adjustment of the two partial-transfer amplitudes, the relative timing of the primitives, the chirped frequency trajectory and the local phase frame. The comparison here supports the practical choice of using the experimentally calibrated sine pulses throughout the main experiment, while showing that finite-window sech pulses can slightly improve the full qutrit-transfer channel.

\subsection{Full-model inverse design of REG waveforms}
\label{sec:supp_ideal_reg_waveform}

The forward simulation discussed in the main text uses inverse-designed REG waveforms. The design is based on the same cascaded input-output model used in the pulse-level simulations and explicitly retains the finite storage amplitudes of both transmission resonators. For either \(e0g1\) or \(f0g1\), we restrict the coherent dynamics to the single-excitation manifold of the addressed transition. Let \(x(t)\) be the sender qutrit amplitude in the active state, \(u(t)\) the sender transmission-resonator amplitude, \(v(t)\) the receiver transmission-resonator amplitude, and \(y(t)\) the receiver qutrit amplitude in the corresponding active state.

The amplitude equations follow directly from the cascaded input-output boundary condition. For a transmission resonator \(j=A,B\) coupled radiatively to the link, the resonant input-output relation is
\begin{equation}
\hat b_{{\rm out},j}(t)=\hat b_{{\rm in},j}(t)+\sqrt{\kappa_j}\,\hat a_j(t),
\label{eq:supp_io_boundary}
\end{equation}
where \(\hat a_j\) is the transmission-resonator annihilation operator. The sender input is vacuum, while the receiver input is the attenuated sender output,
\begin{equation}
\hat b_{{\rm in},B}(t)=\sqrt{\eta_c}\,\hat b_{{\rm out},A}(t),
\label{eq:supp_cascaded_boundary}
\end{equation}
with the propagation phase absorbed into the definition of \(v\) and \(g_B\). We use the standard cascaded-systems quantum-trajectory construction \cite{supp:gardiner1993driving,supp:carmichael1993quantum}. The non-Hermitian Hamiltonian below is the no-jump generator associated with the output and loss channels. In the single-excitation subspace,
\begin{equation}
\hat H_{\rm eff}=\hat H_{\rm sys}+\hat H_{\rm casc}
-\frac{i\hbar}{2}\left(\hat L_{\rm out}^\dagger\hat L_{\rm out}
+\hat L_{\rm loss}^\dagger\hat L_{\rm loss}\right),
\label{eq:supp_quantum_trajectory_heff}
\end{equation}
where
\begin{equation}
\begin{aligned}
\frac{\hat H_{\rm sys}}{\hbar}
&=g_A(t)\left(\hat a_A^\dagger \hat\sigma_A^-+\hat a_A\hat\sigma_A^+\right)
+g_B(t)\left(\hat a_B^\dagger \hat\sigma_B^-+\hat a_B\hat\sigma_B^+\right),\\
\frac{\hat H_{\rm casc}}{\hbar}
&=\frac{i}{2}\sqrt{\eta_c\kappa_A\kappa_B}
\left(\hat a_A^\dagger\hat a_B-\hat a_B^\dagger\hat a_A\right),\\
\hat L_{\rm out}
&=\sqrt{\eta_c\kappa_A}\,\hat a_A+\sqrt{\kappa_B}\,\hat a_B,\qquad
\hat L_{\rm loss}=\sqrt{(1-\eta_c)\kappa_A}\,\hat a_A .
\end{aligned}
\label{eq:supp_cascaded_trajectory_terms}
\end{equation}
Here \(\hat L_{\rm out}\) is the output field after the receiver and \(\hat L_{\rm loss}\) represents channel loss between the two nodes. The no-jump evolution accounts for the conditional amplitude loss into these channels even before the capture pulse is chosen. Expanding Eq.~\eqref{eq:supp_quantum_trajectory_heff} cancels the reciprocal \(\hat a_A^\dagger\hat a_B\) term and leaves only the directional drive \(\hat a_B^\dagger\hat a_A\), giving
\begin{equation}
\begin{aligned}
\frac{\hat H_{\rm eff}}{\hbar}=
&g_A(t)\left(\hat a_A^\dagger \hat\sigma_A^-+\hat a_A\hat\sigma_A^+\right)
+g_B(t)\left(\hat a_B^\dagger \hat\sigma_B^-+\hat a_B\hat\sigma_B^+\right)\\
&-\frac{i\kappa_A}{2}\hat a_A^\dagger\hat a_A
-\frac{i\kappa_B}{2}\hat a_B^\dagger\hat a_B
-i\sqrt{\eta_c\kappa_A\kappa_B}\,\hat a_B^\dagger\hat a_A .
\end{aligned}
\label{eq:supp_cascaded_heff}
\end{equation}
The last term is the unidirectional drive of the receiver resonator by the sender output field; it is the term that would be absent in a bidirectional normal-mode Hamiltonian. In these equations \(\hat\sigma_j^-=\ket{g_j}\bra{s_j}\) is the effective lowering operator of the addressed sideband, with \(s=e\) for \(e0g1\) and \(s=f\) for \(f0g1\). The four basis states in the single-excitation manifold are
\begin{equation}
\begin{aligned}
\ket{x}&=\ket{0_A,s_A;0_B,g_B},&
\ket{u}&=\ket{1_A,g_A;0_B,g_B},\\
\ket{v}&=\ket{0_A,g_A;1_B,g_B},&
\ket{y}&=\ket{0_A,g_A;0_B,s_B},
\end{aligned}
\label{eq:supp_inverse_reg_basis_states}
\end{equation}
where the first and third entries are the sender and receiver transmission-resonator photon numbers, and the second and fourth entries are the sender and receiver qutrit states. Thus \(x\) is the sender qutrit excitation, \(u\) the emitted photon stored in the sender transmission resonator, \(v\) the photon stored in the receiver transmission resonator, and \(y\) the captured receiver qutrit excitation. Applying \(d\ket{\psi}/dt=-i\hat H_{\rm eff}\ket{\psi}/\hbar\) to the no-jump state
\begin{equation}
\ket{\psi(t)}=x(t)\ket{x}+u(t)\ket{u}+v(t)\ket{v}+y(t)\ket{y}
\end{equation}
gives, in the resonant rotating frame,
\begin{equation}
\begin{aligned}
\dot{x}&=-i g_A(t)u,\\
\dot{u}&=-\frac{\kappa_A}{2}u-i g_A(t)x,\\
\dot{v}&=-\frac{\kappa_B}{2}v-\sqrt{\eta_c\kappa_A\kappa_B}\,u-i g_B(t)y,\\
\dot{y}&=-i g_B(t)v ,
\end{aligned}
\label{eq:supp_inverse_reg_amplitudes}
\end{equation}
where \(g_A(t)\) and \(g_B(t)\) are the sender and receiver sideband couplings, \(\kappa_A\) and \(\kappa_B\) are the radiative linewidths of the two transmission resonators, and \(\eta_c\) is the channel efficiency. Perfect absorption imposes the output-field cancellation condition
\begin{equation}
\xi_{\rm out}(t)=\sqrt{\eta_c\kappa_A}\,u(t)+\sqrt{\kappa_B}\,v(t)=0,
\label{eq:supp_inverse_reg_output_cancel}
\end{equation}
This is the dark-output condition: the receiver resonator field destructively interferes with the attenuated sender field so that no outgoing photon remains after the receiver. It is the input-output version of time-reversed photon capture \cite{supp:gardiner1993driving,supp:carmichael1993quantum,supp:korotkov2011flying,supp:wenner2014catching}.

We choose the phase convention \(u(t)=-i q_A(t)\) and \(v(t)=i q_B(t)\), with real mode amplitudes \(q_A\) and \(q_B\). Equation~\eqref{eq:supp_inverse_reg_output_cancel} then gives
\begin{equation}
q_B(t)=\sqrt{\frac{\eta_c\kappa_A}{\kappa_B}}\,q_A(t),
\label{eq:supp_inverse_reg_q_relation}
\end{equation}
and Eq.~\eqref{eq:supp_inverse_reg_amplitudes} becomes
\begin{equation}
\begin{aligned}
\dot{x}&=-g_A q_A,\\
\dot{q}_A&=g_A x-\frac{\kappa_A}{2}q_A,\\
\dot{q}_B&=\sqrt{\eta_c\kappa_A\kappa_B}\,q_A-\frac{\kappa_B}{2}q_B-g_B y,\\
\dot{y}&=g_B q_B .
\end{aligned}
\label{eq:supp_inverse_reg_real_amplitudes}
\end{equation}
Therefore, once a smooth sender-resonator mode \(q_A(t)\) is chosen, the required sideband controls follow analytically:
\begin{equation}
\begin{aligned}
g_A(t)&=\frac{\dot{q}_A+\kappa_A q_A/2}{x(t)},\\
g_B(t)&=\frac{\sqrt{\eta_c\kappa_A\kappa_B}\,q_A-\kappa_B q_B/2-\dot{q}_B}{y(t)} .
\end{aligned}
\label{eq:supp_inverse_reg_controls}
\end{equation}
The remaining amplitudes are fixed by probability conservation in the cascaded no-jump dynamics,
\begin{equation}
\begin{aligned}
x^2(t)&=1-q_A^2(t)-\kappa_A\int_{t_i}^{t}q_A^2(t')\,dt',\\
y^2(t)&=-q_B^2(t)+\int_{t_i}^{t}\left[2\sqrt{\eta_c\kappa_A\kappa_B}\,q_A(t')q_B(t')-\kappa_B q_B^2(t')\right]dt' .
\end{aligned}
\label{eq:supp_inverse_reg_populations}
\end{equation}
This construction is the finite-resonator analogue of choosing an emitted photon mode and solving for the capture pulse that cancels the outgoing field.

For REG, the two primitives are generated in the physical order \(e0g1\) followed by \(f0g1\). A strictly zero-start receiver pulse produces a large initial control when \(y(t)\) is still very small in Eq.~\eqref{eq:supp_inverse_reg_controls}. We therefore use a finite smooth precursor before the main capture window. The sender mode for primitive \(j\in\{e,f\}\) is parameterized as
\begin{equation}
q_{A,j}(t)=A_j \sigma_j^{\alpha_j}(1-\sigma_j)^{\beta_j},\qquad
\sigma_j=\left[1+\exp\left(-\frac{t-t_{c,j}}{\tau_j}\right)\right]^{-1},
\label{eq:supp_inverse_reg_beta_mode}
\end{equation}
over the finite support \(t_{{\rm on},j}<t<t_{{\rm off},j}\), and is zero outside this support. The time \(t_{i,j}\) marks the start of the plotted receiver-capture waveform, not the beginning of the analytic mode: the interval \(t_{{\rm on},j}<t<t_{i,j}\) is a weak precursor that prepares a finite receiver amplitude, and the interval after the nominal capture waveform is retained only to make the turn-off smooth. With this convention \(g_B(t_{i,j})\) can be finite without introducing a sharp initial spike. The normalization is set by the desired partial-transfer probability,
\begin{equation}
\kappa_A\int_{t_{{\rm on},j}}^{t_{{\rm off},j}}q_{A,j}^2(t)\,dt=p_j .
\label{eq:supp_inverse_reg_partial_probability}
\end{equation}
For the lossless maximally entangled REG target, \(p_e=1/3\) for the \(e0g1\) branch and \(p_f=1/2\) for the subsequent \(f0g1\) branch, where \(p_f\) is conditional on the population left after the first partial transfer. The shape parameters and support endpoints can then be optimized within this analytic family, while Eqs.~\eqref{eq:supp_inverse_reg_controls}--\eqref{eq:supp_inverse_reg_populations} provide the corresponding \(g_A(t)\) and \(g_B(t)\) without replacing the model by an effective decay rate.

For qutrit state transfer, pulse optimization targets complete transfer. We therefore use the unwindowed sech pulse \(g(t)=g_0/\cosh[(t-t_0)/\tau]\) with \(g_0/2\pi=1/(2\pi\tau)=5~{\rm MHz}\) and the source-notebook timing convention used in Sec.~\ref{sec:supp_sine_sech_waveform}. Inserted into the full cascaded model with \(\eta_c=0.98\), \(T_1=50~\mu{\rm s}\), Markovian \(T_\phi=20~\mu{\rm s}\) and no additional internal transmission-resonator loss, this qutrit-transfer sequence gives a mean transferred-state fidelity of \(0.968\) and a qutrit process fidelity of \(0.948\).

In the representative REG projection used in the main-text discussion, the waveform start is defined after the weak analytic precursor, so the receiver capture controls begin at finite coupling: \(g_{B,e0g1}(0)/2\pi=2.87~{\rm MHz}\) and \(g_{B,f0g1}(0)/2\pi=3.52~{\rm MHz}\). The full \(200~{\rm ns}\) waveform satisfies \(g_{\rm max}/2\pi=3.95~{\rm MHz}\) and ends with receiver couplings below \(1~{\rm MHz}\). Inserted into the same pulse-level cascaded model with the same projected device parameters, this inverse-designed REG sequence gives \(F=0.939\), \({\cal N}=0.909\), \(C=2.718\) bits and \(I_3=2.688\). These simulated values estimate the performance achievable with the projected device improvements.

\subsection{Simulation error budget for qutrit transfer and REG}
\label{sec:supp_error_budget}

To identify the dominant limitations in the present pulse-level model, we repeated the Fig.~3 qutrit-transfer and Fig.~4 REG simulations while adding the calibrated error channels one at a time. The first row of Table~\ref{tab:supp_qst_reg_error_budget} keeps the experimentally used pulse sequence and pulse amplitudes, but removes photon loss, internal resonator loss, qutrit relaxation and dephasing. This row therefore includes coherent-control imperfections of the calibrated sine-pulse sequence. The following rows add, in order, finite channel efficiency \(\eta_c\), internal transmission-resonator loss \(\kappa_{\rm int}\), measured qutrit \(T_1\), and the quasi-static \(1/f\)-like dephasing ensemble. The resulting error budget is cumulative, with each error channel fixed by its calibration.

\begin{figure}[htbp]
    \centering
    \suppfiginclude{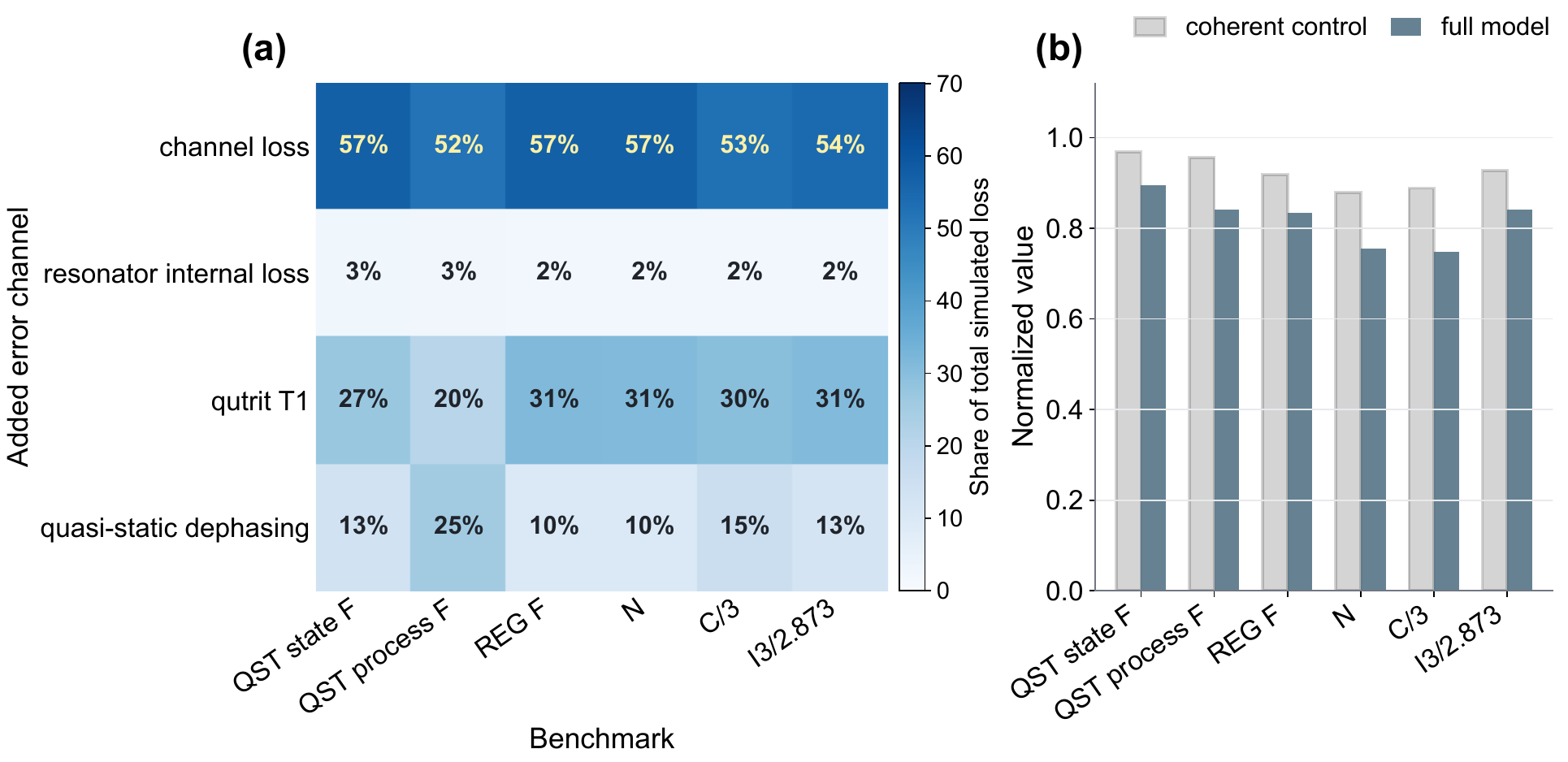}{width=0.95\textwidth}
    \caption{\textbf{Simulation error budget for qutrit transfer and REG.}
    \textbf{(a)} Incremental contribution of each added error channel to the total simulated loss between the coherent-control row and the full-model row in Table~\ref{tab:supp_qst_reg_error_budget}.
    \textbf{(b)} Normalized coherent-control and full-model values for the same benchmarks. Capacity is normalized by 3 bits and \(I_3\) by the ideal qutrit CGLMP value \(2.873\). For the REG nonlinear metrics, the error budget uses the same averaged-density-matrix convention as the main-text Fig.~4 simulation: the quasi-static density matrices are averaged before evaluating \({\cal N}\), \(C\), and \(I_3\).}
    \label{fig:supp_qst_reg_error_budget}
\end{figure}

\begin{table}[htbp]
\centering
\caption{\textbf{Simulation error budget for qutrit transfer and REG.}
The first line is the coherent-control result obtained with the calibrated pulse sequence but with \(\eta_c=1\), \(\kappa_{\rm int}=0\), no qutrit relaxation and no dephasing. Subsequent lines add the physical error channels cumulatively.}
\label{tab:supp_qst_reg_error_budget}
\begin{tabular}{lcccccc}
\hline
Model & QST state F & QST process F & REG F & \({\cal N}\) & \(C\) (bits) & \(I_3\) \\
\hline
Coherent control only & 0.969 & 0.956 & 0.918 & 0.879 & 2.667 & 2.666 \\
\(+\) channel loss & 0.927 & 0.896 & 0.870 & 0.808 & 2.443 & 2.531 \\
\(+\) resonator internal loss & 0.925 & 0.893 & 0.868 & 0.806 & 2.434 & 2.526 \\
\(+\) qutrit \(T_1\) & 0.905 & 0.870 & 0.842 & 0.768 & 2.308 & 2.448 \\
\(+\) quasi-static dephasing & 0.895 & 0.841 & 0.834 & 0.756 & 2.242 & 2.417 \\
\hline
\end{tabular}
\end{table}

The budget shows two different regimes. For qutrit state transfer, finite channel efficiency gives the largest single drop in both the mean transferred-state fidelity and the process fidelity. The measured qutrit \(T_1\) is the next largest contribution, while internal resonator loss is small at the present \(\kappa_{\rm int}/2\pi\) values. The quasi-static dephasing ensemble has a modest effect on the average state fidelity, but a larger effect on the process fidelity because the process matrix is sensitive to coherent phase errors across the full input-state set.

For REG, the dominant simulated limitations in this convention are finite channel efficiency and qutrit relaxation, while quasi-static dephasing gives a smaller but still phase-sensitive reduction. For the nonlinear entanglement and information metrics in this diagnostic table, we use the same averaged-density-matrix convention as in the main-text Fig.~4 comparison: the density matrices from the quasi-static detuning ensemble are averaged before \({\cal N}\), \(C\), and \(I_3\) are evaluated. This shot-mixed convention gives a stricter estimate of how intra-block phase wandering would reduce the reconstructed two-qutrit resource than averaging nonlinear metrics over independently phase-stable trajectories. In this limit, quasi-static dephasing reduces the negativity, dense-coding capacity and \(I_3\), which depend on the phase-sensitive interference terms in the logical qutrit Bell basis. This analysis supports the device-improvement priorities inferred from the data: increasing the channel efficiency, extending qutrit \(T_1\), reducing low-frequency phase noise, and replacing the experimental sine pulses by REG waveforms designed in the full cascaded model should provide the largest immediate gains, while lower internal transmission-resonator loss gives a smaller but still systematic improvement.

\clearpage
\makeatletter\let\pre@bibdata\@empty\makeatother
\end{document}